\documentclass[12pt]{article}
\usepackage[english]{babel}
\usepackage[pdftex]{color}
\usepackage{amsmath}
\usepackage{graphicx}
\usepackage{hyperref}
\usepackage{amsmath}
\usepackage{amsfonts}
\usepackage{amssymb}

\begin{document}

\title{ \textbf{{}BRST-BV  Quantum  Actions for Constrained Totally-Symmetric Integer HS Fields}}
\author{\textsc{\v{C}estmir  Burd\'{\i}k${}^{a}$\thanks{burdik@kmlinux.fjfi.cvut.cz  \hspace{0.5cm} ${}^{\dagger}$reshet@tspu.edu.ru}, Alexander A. Reshetnyak${}^{b,c,d\dagger}$}\\
 \it ${}^a$Department of Mathematics, Czech Technical University,\\
 \it  Prague  12000,  Czech
Republic,\\[0.2cm]
\it ${}^{b}$Tomsk State Pedagogical University, Tomsk 634041,   Russia,\\[0.2cm]
\it ${}^{c}$National Research Tomsk State University, Tomsk 634050,  Russia,\\[0.2cm]
\it ${}^{d}$Institute of
Strength Physics and Materials Science SB RAS,\\ \it Tomsk 634055,  Russia}
\date{}
\maketitle
\begin{abstract}
 A constrained BRST--BV Lagrangian formulation for totally
symmetric massless HS fields in a $d$-dimensional Minkowski space
is extended to a non-minimal constrained BRST--BV Lagrangian
formulation by using a non-minimal BRST operator
$Q_{c|\mathrm{tot}}$ with non-minimal Hamiltonian BFV oscillators
$\overline{C}, \overline{\mathcal{P}}, \lambda, \pi$, as well as
antighost and Nakanishi-Lautrup tensor fields, in order to
introduce an admissible self-consistent gauge condition. The
gauge-fixing procedure involves an operator gauge-fixing BRST-BFV
Fermion $\Psi_H$ as a kernel of the gauge-fixing BRST--BV Fermion
functional $\Psi$, manifesting the concept of BFV--BV duality. A
Fock-space quantum action with non-minimal BRST-extended off-shell
constraints is constructed as a shift of the total generalized
field-antifield vector by a variational derivative of the
gauge-fixing Fermion $\Psi$ in a total BRST--BV action  $
S^{\Psi}_{0|s}  =     \int d \eta_0 \langle \chi^{\Psi{}
0}_{\mathrm{tot}|c} \big| Q_{c|\mathrm{tot}}\big| \chi^{\Psi{}
0}_{\mathrm{tot}|c}\rangle$. We use a gauge condition which
depends on two gauge parameters, thereby extending the case of
$R_\xi$-gauges. For triplet and doublet formulations we explored the representations with only traceless field-antifield and source variables.  For the generating functionals of Green's
functions, BRST symmetry transformations are suggested and Ward
identities are obtained.
\end{abstract}

\thispagestyle{empty}
\emph{The paper is dedicated to the loving memory of the aunt and
 godmother of A.A.R., Lyubov' Fyodorovna Dyachenko.}

\section{Introduction}

Many of the topical issues in high-energy physics are related to
higher-spin (HS) field theory as part of the LHC experiment
program. Various tensionless limits of (super)string theory
\cite{tensionlessl} implied by their respective BRST operators
contain an infinite set of HS fields with integer and half-integer
generalized spins, as well as a set of HS fields with continuous
generalized spin \cite{Savvidy}, \cite{Mourad} (for another
viewpoint, see  \cite{1302.4771}). This incorporates HS field
theory into superstring theory and transforms it into a method for
studying the classical and quantum structures of the latter (for
the present status of HS field theory, see the reviews
\cite{reviews}, \cite{reviewsV}, \cite{reviews3},
\cite{Vasiliev5}). One of the most efficient ways of studying HS
field dynamics, including Lagrangian formulations (LF) in constant
curvature spaces and starting from the initial unitary irreducible
representation (UIR) of the Poincar\'{e} or anti-de-Sitter groups,
is based on a constrained BRST--BFV approach for lower integer
spins, examined for the first time in  \cite{BRST-BFV1}, \cite{firstBRST}  and originated from the so-called minimal part of
the BFV method \cite{BFV}, \cite{BFV1}, devised for quantizing
constrained dynamical systems. The BRST--BFV approach is intended
to solve the inverse problem, which is formulating an LF in terms
of Hamiltonian-like objects by using an auxiliary Hilbert space
whose vectors consist of HS (spin)-tensor fields. Incorporating
holonomic (traceless and mixed-symmetry) constraints, together
with differential ones, into a total system of constraints
(without any restrictions imposed on the entire set of the initial
and auxiliary HS fields) -- which is to be closed with respect to
Hermitian conjugation supplied with an appropriate conversion
procedure for the subsystem of second-class constraints -- has
resulted in augmenting the original method by an unconstrained
BRST--BFV method. Applications of this method have been originated
by \cite{Pashnev1} and followed by \cite{Pashnev2},
\cite{BuchPashnev}, \cite{totfermiMin}, \cite{0505092},
\cite{totfermiAdS} for totally-symmetric HS fields and
mixed-(anti)symmetric HS fields in $R^{1,d-1}$ and AdS${}_d$
\cite{Pashnev3}, \cite{ReshMosh}, \cite{BurdikResh},
\cite{BuchbResh}, \cite{Reshetnyk2}, \cite{Reshetnyak_mas}; for a
review and the interaction problem, see \cite{reviews3}. A
detailed correspondence between the constrained and unconstrained
BRST--BFV methods for arbitrary massless and massive IR of the
$ISO(1,d-1)$ group with a generalized discrete spin has been
recently studied in \cite{Reshetnyak_con}, where equivalence
between the underlying constrained and unconstrained LF was
established. A development of this topic has resulted in an
(un)constrained BRST--BV method of finding minimal BV actions
required to construct a quantum action in the BV quantization
\cite{BV} presented by \cite{ReshBRSTBV}; for bosonic HS fields,
see also \cite{BRST-BV1}, \cite{BRST-BV2}, \cite{BRST-BV3}.
Recently, the issues of LF construction and dynamics for
continuous spin particles (CSP) have been analyzed in the
Shuster--Toro representation for bosonic \cite{ShusterToro} and fermionic  \cite{bekaertNajafizadeh} fields by R.~Metsaev
\cite{Metsaevcontbos}  and also in \cite{BuchKrychcont} using the
Weyl spinor notation -- for recent developments, see also
\cite{Metsaev2}, \cite{Najafizadeh}, \cite{bekaertSkvortsov}, \cite{Zinoviev},
\cite{Alkalaev}, \cite{FedorukBuch2} -- whereas constrained
BRST--BFV and BRST--BV descriptions of CSP particle dynamics using
the original Wigner--Bargmann equations \cite{Wigner1} have been
examined in \cite{BPR}, along with a special tensionless limit of
string theory.

The quantization problem lies in constructing a so-called
Batalin--Vilkovisky action for (general) reducible gauge-invariant
LF. Until now, the problem has been solved by representing
BRST-BFV or BRST--BV Lagrangians in component or oscillator forms
without Hamiltonian operator oscillators in the minimal sector
\cite{BFV}, \cite{BFV1}, following the BV method \cite{BV} (or its
simplified version of the Faddeev--Popov recipe \cite{FP} for
irreducible theories with closed algebras), as was done, for
instance, in \cite{MetsaevQA}; see also \cite{Siegel}. At the same
time (because the set of all monomials in the powers of minimal
Hamiltonian oscillators is in one-to-one correspondence with the
set of all field and antifield vectors) the set of non-minimal
Hamiltonian oscillators of the BFV method, as well as the set of
antighost $|\overline{C}\rangle$ and Nakanishi--Lautrup
$|b\rangle$ field vectors from the non-minimal sector of the BV
field-antifield space, has not been utilized to form the field
$|\overline{C}\rangle$, $|b\rangle$ and the related antifield
vectors. This fact has prevented one from imposing admissible
gauge-fixing condition as a respective shift of antifields in the
BV action. Having in mind an equivalence between unconstrained and
constrained BRST--BFV LF for one and the same HS field of
generalized discrete spin in $\mathbb{R}^{1,d-1}$
\cite{Reshetnyak_con}, we restrict ourselves by a massless
totally-symmetric (TS) HS field with integer helicities $s\in
\mathbb{N}_0$ in a constrained BRST-BFV LF. The paper is devoted
to the following problems:
\begin{enumerate}
\item  Enlargement of constrained BRST--BV Lagrangians for TS
massless fields with integer helicity in a $d$-dimensional
Minkowski space-time to  non-minimal constrained BRST--BV
Lagrangians with a compatible set of off-shell BRST-extended
constraints in the metric formulation;

\item  Introduction of a gauge-fixing procedure in order to
eliminate the gauge degeneracy of a classical BRST--BFV action so
as to construct a quantum (non-renormalized) BRST--BV action and a
generating functional of Green's functions with an essential
self-consistent use of variables from non-minimal sectors
originated from both the operator BFV and functional BV methods.
\end{enumerate}

The paper is organized as follows. In Section~\ref{cBRSTBFV}, we
overview the constrained BRST--BV and BRST--BFV approaches in the
minimal sector of field-antifield variables for a TS HS field of
integer helicity. The construction of a total BV action on the
basis of a constrained BRST--BV approach augmented by respective
non-minimal field-antifield tensor fields and non-minimal
Hamiltonian oscillators for the TS HS fields is considered in
Section~\ref{nonBRSTBV}. In Section~\ref{fgBRSTBV}, we present a
gauge-fixing procedure and representations (in case of doublet, triplet formulations) for the initial field in terms of sum of 2 traceless fields used to construct quantum BV actions with
an underlying BRST symmetry for respective generating functionals
of Green's functions, and make concluding comments in
Section~\ref{Concl}.

We use the convention $\eta_{\mu\nu} = diag (+, -,...,-)$ for the
metric tensor, with Lorentz indices $\mu, \nu = 0,1,...,d-1$, and
the respective notation $\epsilon(F)$, $[gh_{H},gh_{L},
gh_{tot}](F)$ for the value of Grassmann parity and those of the
BFV, $gh_{H}$, BV, $gh_{L}$ and total $gh_{tot}=gh_{H}+gh_{L}$
ghost number of a homogeneous quantity $F$. The supercommutator
$[F,\,G\}$ of any quantities $F$ and $G$ with definite values of
Grassmann parity is given by $[F\,,G\}$ = $FG
-(-1)^{\epsilon(F)\epsilon(G)}GF$.

\section{Constrained BRST-BFV and BRST-BV Lagrangians }

\label{cBRSTBFV}

Recall that the UIR of $ISO(1,d-1)$ group  with of zero mass and
integer helicity $s\in \mathbb{N}_0$ is realized using an
$\mathbb{R}$-valued TS tensor field $\phi_{\mu_1...\mu_s}(x)\equiv
\phi_{(\mu)_s}$ and described by the following equivalent conditions
\begin{equation}\label{irrepint}
     \big(\partial^\nu\partial_\nu,\, \partial^{\mu_1},\, \eta^{\mu_1\mu_2}\big)\phi_{(\mu)_s}  = (0,0,0)   \  \Longleftrightarrow  \  \big(l_0,\, l_1,\, l_{11}, g_0 -d/2\big)|\phi\rangle  = (0,0,0,s)|\phi\rangle
\end{equation}
for a basic vector and operators in a  Fock space
$\mathcal{H}$  generated by  Grassmann-even (symmetric basis)
oscillators $a_\mu, a^+_\nu$ ($[a_\mu, a^+_\nu] = -
\eta_{\mu\nu}$)
\begin{equation}\label{FVoper}
  |\phi\rangle  =  \sum_{s\geq 0}\frac{\imath^s}{{s!}}\phi^{(\mu)_s}\prod_{i=1}^s a^+_{\mu_i}|0\rangle, \ \  \big(l_0,\, l_1,\, l_{11}, g_0\big) = \big(\partial^\nu\partial_\nu,\, - \imath a^\nu  \partial_\nu ,\, \frac{1}{2}a^\mu a_\mu ,  -\frac{1}{2}\big\{a^+_{\mu}.\, a^{\mu}\big\}\big).
\end{equation}
The constrained BRST-BFV approach for a free TS massless HS field
$\phi^{(\mu)_s}$ in Minkowski space results in an irreducible
gauge-invariant LF with a nilpotent hermitian constrained BRST
operator $Q_c$, an off-shell BRST-extended  traceless constraint
$\widehat{L}_{11}$ and a spin operator $\widehat{\sigma}_c(g)$
acting in a total Hilbert space $\mathcal{H}{}_{c|tot}=
\mathcal{H}\otimes H^{o_a}_{gh}$ with scalar product
$\langle\bullet| \bullet\rangle$, which admits a
$\mathbb{Z}$-grading $\mathcal{H}{}_{c|tot} = \oplus_{e}
\mathcal{H}{}^e_{c|tot}$ corresponding to  the ghost number
$gh_H$, $gh_H(\mathcal{H}{}^e_{c|tot})=-e$:
\begin{eqnarray}
\hspace{-0.5ex}&\hspace{-0.5ex}&\hspace{-0.5ex}\label{PhysStatetot} \mathcal{S}_{C|s}(\phi,\phi_1,\phi_2)= \int d\eta_0 {}_s\langle\chi^0_c| Q_c|\chi^0_c\rangle_s, \ \delta |\chi^0_c\rangle_s = Q_c|\chi^1_c\rangle_s ,   \ \ (\epsilon, gh_H) |\chi^e_c\rangle = (e, -e),
\\
\hspace{-0.5ex}&\hspace{-0.5ex}&\hspace{-0.5ex} {Q_c} =  \eta_0l_0+\eta_1^+l_1+l_1^{+}\eta_1
 + {\imath}\eta_1^+\eta_1{\cal{}P}_0 \ = \ \eta_0l_0+\Delta Q_c  + {\imath}\eta_1^+\eta_1{\cal{}P}_0  ,  \ \ (\epsilon, gh_H) Q_c = (1, 1),  \label{Qctotsym}
  \\
 \hspace{-0.5ex}&\hspace{-0.5ex}&\hspace{-0.5ex} \big(\widehat{L}_{11},\, \widehat{\sigma}_c\big) |\chi^e_c\rangle_s =\Big(0,\,s+\frac{d-2}{2}\Big)|\chi^e_c\rangle_s, \ \ \big(\widehat{L}_{11} ,\,\widehat{\sigma}_c \big) =  \big( l_{11}+\eta_{1} \mathcal{P}_{1} , \,  g_0+ \eta_1^+\mathcal{P}_{1} -\eta_1\mathcal{P}_{1}^+ \big) \label{extconstsp} \end{eqnarray}
  for $e=0,1$ and $ |\chi^e_c\rangle_s \in \mathcal{H}{}^e_{c|tot}$.
  Here, the subspace $H^{o_a}_{gh}$  is generated
  by some additional (to $a_{\mu}, a^{+}_{\mu}$)  BFV
  Grassmann-odd ghost operators from the  minimal
  sector $\{C^a, \mathcal{P}_a\}$ = $\{\eta_0, \mathcal{P}_0$;
  $\eta_1, \mathcal{P}_1^+$; $\eta^+_1, \mathcal{P}_1\}$,
  with the ghost number distribution $gh_H (\eta) $ =$-gh_H
  (\mathcal{P})$ = $1$ and the non-vanishing anticommutators
  $\{\eta_0, \mathcal{P}_0\}= \imath$, $ \{\eta_1, \mathcal{P}_1^+\}=1$.
  These ghost operators  are introduced for the system
  of  first-class differential constraints
  $\{o_a\}$ =  $\{l_0, l_1, l_1^+ \}$: $l_1^+ = - \imath a^{+\nu}
  \partial_\nu$ subject to the algebra:
  $[l_0, l^{(+)}_1] = 0$, $[l_1,l_1^+]=l_0$.
  The operators $Q_c, \widehat{L}_{11},\, \widehat{\sigma}_c$
  are found  as solutions of the generating equations
  \cite{Reshetnyak_con}
  \begin{equation}\label{geneq}
   Q_c^2 = 0,\ \ \  [Q_c,\, \widehat{L}_{11}\} = 0, \ \ \  [Q_c,\, \widehat{\sigma}_c\} =0,\ \ \   [\widehat{L}_{11},\, \widehat{\sigma}_c\} =  2\widehat{L}_{11}.
  \end{equation}
The field $ |\chi^0_c\rangle_s$ and the gauge parameter
$|\chi^1_c\rangle_s$ labelled by the symbol $"s"$ as eigenvectors
of the spin condition in  (\ref{extconstsp}) read as follows ($\phi_2(a^+)\rangle \equiv 0$ when  $s\leq 1$ and $\phi_1(a^+)\rangle \equiv 0$ for  $s = 0$):
\begin{eqnarray}
\hspace{-1em}&\hspace{-1em}&\hspace{-1em} |\chi^0_c\rangle_s  = |S_c\rangle_s+\eta_0|B_c\rangle_s = |\phi(a^+)\rangle_s+\eta_1^+\mathcal{P}_1^+|\phi_2(a^+)\rangle_{s-2}+\eta_0\mathcal{P}_1^+|\phi_1(a^+)\rangle_{s-1} \label{spinctotsym} \\
\hspace{-1em}&\hspace{-1em}&\hspace{-1em} \phantom{ |\chi^0_c\rangle_s} =  \Big( \hspace{-0.1em} \frac{\imath^s}{{s!}}\phi^{(\mu)_s}\hspace{-0.1em}\prod_{i=1}^s\hspace{-0.1em} a^+_{\mu_i} +\eta_1^+\mathcal{P}_1^+ \frac{\imath^{s-2}}{{(s-2)!}}\phi_2^{(\mu)_{s-2}}\hspace{-0.1em}\prod_{i=1}^{s-2}\hspace{-0.1em} a^+_{\mu_i} +\eta_0\mathcal{P}_1^+ \frac{\imath^{s-1}}{{(s-1)!}}\phi_1^{(\mu)_{s-1}}\hspace{-0.1em}\prod_{i=1}^{s-1}\hspace{-0.1em} a^+_{\mu_i} \hspace{-0.1em} \Big)\hspace{-0.1em} |0\rangle , \label{spinctotsym2}\\
\hspace{-1em}&\hspace{-1em}&\hspace{-1em} |\chi^1_c\rangle_s  = \mathcal{P}_1^+  |\xi(a^+)\rangle_{s-1} =  \mathcal{P}_1^+ \frac{\imath^{s-1}}{{(s-1)!}}\xi_{(\mu)_{s-1}}\prod_{i=1}^{s-1}a^{+\mu_{i}}|0\rangle.
  \label{parctotsym}
\end{eqnarray}
Solving the traceless constraints (\ref{extconstsp}) leads to
the following relations for the fields:
\begin{eqnarray}
    && l_{11}\left(|{\phi}\rangle; |{\phi}_1\rangle, |{\phi_2}\rangle , |\xi\rangle\right) = (-|{\phi_2}\rangle; 0,0,0 ) \  \Longleftrightarrow \  \nonumber\\
    && \left(\phi^{(\mu)_{s-2}\mu}{}_{\mu}; \phi_1^{(\mu)_{s-3}\mu}{}_{\mu}, \phi_2^{(\mu)_{s-4}\mu}{}_{\mu}, \xi^{(\mu)_{s-3}\mu}{}_{\mu}\right)=\left(2\phi_2^{(\mu)_{s-2}};0,0,0\right). \label{resconstrf}
\end{eqnarray}
The gauge-invariant  action $\mathcal{S}_{C|s}=\mathcal{S}_{C|s}(\phi,\phi_1,\phi_2)$ is written in the triplet: $\eta_0-$independent,   ghost-independent and tensor   forms \cite{FranciaSagnottitrip}
\begin{eqnarray}
 &&  \label{Sclsr} \mathcal{S}_{C|s}  =  \left(\hspace{-0.2em}{}_{s}\langle S_c \big|   {}_{s}\langle B_c\big|  \hspace{-0.2em}\right)\left(\hspace{-0.2em}\begin{array}{cc}
  l_0 & - \Delta Q_c \\
 -\Delta Q_c & \eta_1^+\eta_1            \end{array}\hspace{-0.2em}\right)
       \left( \hspace{-0.2em} \begin{array}{c}\big|{S}_c\rangle_s \\ \big|{B}_c \rangle_s  \end{array}\hspace{-0.2em} \right),  \delta \left( \hspace{-0.2em} \begin{array}{c}\big|{S}_c\rangle_s \\ \big|{B}_c \rangle_s  \end{array} \hspace{-0.2em}\right) = \left(\hspace{-0.2em}\begin{array}{c}
\Delta Q_c  \\
           l_0  \end{array}\hspace{-0.2em}\right)
       \big|{S}^{1}\rangle_s , \\
       && \mathcal{S}_{C|s}  =  \left({}_{s}\langle \phi  \big|    {}_{s-2}\langle \phi_2\big| {}_{s-1}\langle \phi_1\big|   \right)\left(\begin{array}{ccc}
  l_0 &   0 & -l_1^+ \\
 0 & -l_0 &  l_1    \\
        -l_1 & l_1^+&  1  \end{array}\right)
       \left(  \begin{array}{c}\big|{\phi}\rangle_s\\ \big|{\phi_2}\rangle_{s-2} \\ \big|{\phi_1}\rangle_{s-1}   \end{array} \right) ,  \label{gaugetrip}\\
                \label{gaugetr}
                &&
    \phantom{\mathcal{S}_{C|s}} \delta \left( \big|\phi\rangle_{s} ,      \big|\phi_1\rangle_{s-1},    \big|\phi_2\rangle_{s-2}\right) = \left( l_1^+ ,l_0,l_1\right) |\xi\rangle_{s-1} \, , \\
    \hspace{-0.5em} &\hspace{-0.5em}&\hspace{-0.5em}   \mathcal{S}_{C|s}  =  \frac{(-1)^s}{s!} \int d^dx\bigg\{\phi_{(\mu)_s}\big(\partial^2\phi^{(\mu)_s} +2{s} \partial^{\mu_s} \phi_1^{(\mu)_{s-1}}\big) -
   {s(s-1)}\phi_{2(\mu)_{s-2}}\partial^2\phi_2^{(\mu)_{s-2}}  \label{tenstripl}\\
  \hspace{-1em}&\hspace{-1em}&\hspace{-1em}\phantom{ \mathcal{S}_{C|s}} \   - s\phi_{1(\mu)_{s-1}}\big( \phi_1^{(\mu)_{s-1}}  - 2(s-1) \partial^{\mu_{s-1}} \phi_2^{(\mu)_{s-2}}\big)  \bigg\} ,   \nonumber \\
&&  \phantom{\mathcal{S}_{C|s}} \delta\big(\phi^{(\mu)_{s}}, \phi_1^{(\mu)_{s-1}}, \phi_2^{(\mu)_{s-2}}\big)\ =\ \big( -\partial^{(\mu_s}\xi^{(\mu)_{s-1})}, \, \partial^2\xi^{(\mu)_{s-1}},\, \partial_{\mu_{s-1}} \xi^{(\mu)_{s-1}} \big)\nonumber
\end{eqnarray}
(for $|{S}^{1}\rangle_s  \equiv |{\chi}^{1}_c\rangle_s$) as well as in the doublet  form with $\mathcal{S}^d_{C|s}=\mathcal{S}_{C|s}|_{\big(\phi_1=\phi_1(\phi,\phi_2)\big)}$ (having expressed $\big|\phi_1\rangle_{s-1}$ from the equation of motion: $\big|\phi_1\rangle = l_1\big|\phi\rangle- l^+_1\big|\phi_2\rangle $)
\begin{eqnarray}\label{dublet}
       && \mathcal{S}^d_{C|s}  =  \left({}_{s}\langle \phi  \big|    {}_{s-2}\langle \phi_2\big|    \right)\left(\begin{array}{cc}
  l_0 -l_1^+l_1&    (l_1^+)^2 \\
 l^2_1 & -l_0-l_1l_1^+      \end{array}\right)
       \left(  \begin{array}{c}\big|{\phi}\rangle_s\\ \big|{\phi_2}\rangle_{s-2}    \end{array} \right) ,
       \end{eqnarray}
       and in the single field form  (Fronsdal \cite{Fronsdal}):
       \begin{eqnarray}
 &&  \label{SclsrsingleF}  \mathcal{S}_{F|s}\left({\phi}\right)  =  {}_{s}\langle \phi  \big|    \left(  l_0-l_1^+l_1  -(l_1^+)^2l_{11}
 -l_{11}^+l_1^2  -l_{11}^+(l_0 +  l_1l_1^+) l_{11}    \right)
       \big|{\phi}\rangle_s, \end{eqnarray}
        \vspace{-1ex}\begin{eqnarray}    &&     \delta  \big|\phi\rangle_{s}  =  l_1^+  |\xi\rangle_{s-1}\ \   \mathrm{and}\ \
            l_{11}\big(l_{11}|{\phi}\rangle,\,  |\xi\rangle\big) = (0,0 ) ,\label{gaugetrsing}\\
   \hspace{-1.2em}&\hspace{-1.2em}&\hspace{-1.2em} \mathcal{S}_{F|s}(\phi) \hspace{-0.1em}  = \hspace{-0.1em} \frac{(-1)^s}{s!} \int d^dx\bigg\{\phi_{(\mu)_s}\big(\partial^2\phi^{(\mu)_s}  -{s} \partial^{\mu_s} \partial_{\nu}\phi^{(\mu)_{s-1}\nu} +  {s}(s-1)\partial^{\mu_{s-1}}\partial^{\mu_s} \phi^{(\mu)_{s-2}\mu}{}_{\mu}\big)  \nonumber\\
  \hspace{-1em}&\hspace{-1em}&\hspace{-1em} \  -
   \frac{1}{2}{s(s-1)}\phi_{(\mu)_{s-2}\mu}{}^{\mu}\big(\partial^2\phi^{(\mu)_{s-2}\nu}{}_{\nu} + \frac{1}{2} (s-2)\partial^{\mu_{s-2}}\partial^{\mu}\phi^{(\mu)_{s-3}}{}_{\mu\nu}{}^{\nu}  \big) \bigg\}.   \label{Fronsdal1}\end{eqnarray}

In its turn, the constrained BRST--BV method of finding a BV
action in the minimal sector of field-antifield variables augments
the BRST--BFV algorithm by transforming the gauge parameter
$|\chi^1_c \rangle$ into a  ghost  field $|C \rangle$, thereby
introducing the respective antifields $|\chi^{*0}_c \rangle$,
$|C^* \rangle$ and incorporating a unique generalized
field-antifield vector.

Depending on a given BRST--BFV LF (triplet, doublet, or Fronsdal),
the extension of the configuration space
$\mathcal{M}_{\mathrm{cl}}$, due to other HS tensors, up to a
minimal configuration space $\mathcal{M}_{\mathrm{min}}$ in the
case of a TS field is determined by a \emph{generalized vector}
$|\chi_{\mathrm{g}|c} \rangle$ from a generalized Hilbert space
$\mathcal{H}{}_{g|tot}= \mathcal{H}_g\otimes H^{o_a}_{gh}$
(instead of $|\chi_c \rangle$ $\in \mathcal{H}{}_{c|tot}$),
\vspace{-1.5ex}\begin{eqnarray}
|\chi_{\mathrm{g}|c} \rangle &=&  \sum_{n}\frac{\imath^{n}}{ n!} \Big(\prod ( \eta_0
)^{n_{f 0}}  \prod( \eta_1^+ )^{n_{f }} (
\mathcal{P}_1^+ )^{n_{p }}   \phi^{n_{f 0} n_{f }n_{p }}_{\mathrm{g}|c{} (\mu)_{n}} \prod_{i=1}^n  a^{\mu_i+} \Big)|0\rangle,  \label{chirgen}
\end{eqnarray}
for $ |\phi(a^+)^{n_{f 0} n_{f } n_{p }}_{\mathrm{g}|c }\rangle \equiv \phi^{n_{f 0} n_{f}n_{p }}_{\mathrm{g}|c{} (\mu)_{n}} \prod_{i=1}^n  a^{\mu_i+} |0\rangle$. Here any ghost independent vector  $\phi(a^+)^{\ldots}_{\mathrm{g}|c}|0\rangle  \in  \mathcal{H}_g$  with a vanishing BV ghost number  [$gh_L(\phi(a^+)^{\ldots}_{\mathrm{g}|c{}}) =0$]  coincides with the field vector $\phi(a^+)^{\ldots}_{c} |0\rangle  \in  \mathcal{H}_c$ for the vector $|\chi_c \rangle$ also determined by the  representation (\ref{chirgen}).
The space $\mathcal{H}{}_{g|tot}$ admits a $\mathbb{Z}\oplus \mathbb{Z}$-grading  corresponding  to  the ghost numbers $gh_H, gh_L$ distributions
\begin{equation}\label{Zgradtot}
 \mathcal{H}{}_{g|tot} = \oplus_{e,l} \mathcal{H}{}^{e,l}_{g|tot}: \quad  gh_H(\mathcal{H}{}^{e,l}_{g|tot})=-e,\  gh_L(\mathcal{H}{}^{e,l}_{g|tot})=l
\end{equation}
 From the same spectral problem for $Q_c$-complex  with imposing of spin and  BRST-extended constraint in $\mathcal{H}{}_{g|tot}$:
 \begin{equation}\label{BRSTcomplex}
   Q_c|\chi^0_{\mathrm{g}|c}\rangle  =0, \  \ \delta |\chi^l_{\mathrm{g}|c}\rangle = Q_c|\chi^{l+1}_{\mathrm{g}|c}\rangle,  \quad   \big(\widehat{L}_{11},\, \widehat{\sigma}_c\big) |\chi^l_{\mathrm{g}|c}\rangle =\Big(0,\,s+\frac{d-2}{2}\Big)|\chi^l_{\mathrm{g}|c}\rangle
 \end{equation}
albeit with a vanishing total ghost number for  $|\chi^0_{\mathrm{g}|c}\rangle$, (i.e. without using the sequence of gauge transformations: $\delta |\chi^l_{\mathrm{g}|c}\rangle = Q_c|\chi^{l+1}_{\mathrm{g}|c}\rangle  =0, $ for $l=0,1,...$)     we obtain the spin and
modified ghost numbers distributions for  proper eigen-vector $|\chi^0_{\mathrm{g}|c} \rangle_s$: $gh_{tot}(|\chi^0_{\mathrm{g}|c} \rangle_s) = 0$.  This means, that  $|\chi^0_{\mathrm{g}|c} \rangle \in  \mathcal{H}{}^{0}_{g|tot}$ for  $\mathcal{H}{}^{p}_{g|tot} \equiv $ $\oplus_{e+l=p} \mathcal{H}{}^{e,l}_{g|tot}$.
 The whole set of fields $\phi^A_{\min}$  and antifields $\phi^*_{A\min}$  [$\big(\epsilon, gh_L\big)\phi^*_{A\min}$ = $\big(\epsilon(\phi^A_{\min})+1, - gh_L(\phi^A_{\min})-1\big)$] from the  minimal BV sector  for given BRST-BFV (triplet)  LF with the use of the condensed notations:
\begin{eqnarray}\label{confspmin}
&& \{\phi^A_{\min}\}  = \big\{ \phi_k^{(\mu)_{s-k}},   C^{(\mu)_{s-1}}\big\}(x) , \ \  \{\phi^*_{A\min}\} = \big\{ \phi^*_{k(\mu)_{s-k}}, C^*_{(\mu)_{s-1}}\big\}(x),
\end{eqnarray}
is in one-to-one correspondence with the set of  field components   in the \emph{minimal generalized  vector}  of spin $s$
\begin{eqnarray}\label{genfstr}
 && |\chi^0_{\mathrm{g}|c}\rangle_s \ =\  |\chi^0_{c}\rangle_s +  \mathcal{P}_1^+C_{s-1}(a^+)|0\rangle + |\chi^{0\ast}_{c}\rangle_s - \eta_0\eta_1^+C^*_{s-1}(a^+)|0\rangle.\\
 \hspace{-0.5em}&\hspace{-0.5em}&  \hspace{-0.5em}|\chi^{0\ast}_c\rangle_s  \  = \    |S^* _c\rangle_s+ \eta_0|B^*_c\rangle_s \ =\   \eta_1^+ |\phi^*_1(a^+)\rangle_{s-1}+ \eta_0 |\phi^*(a^+)\rangle_s+ \eta_0\mathcal{P}_1^+\eta^+_1|\phi^*_2(a^+)\rangle_{s-2}.\label{genfstr1}
\end{eqnarray}
 The properties of $\mathbb{Z}$  gradings for all BRST-BV
oscillators and (anti)field variables are presented in
Table~\ref{table gh}.
 \begin{table}[t]
\begin{center}\vspace{-2.5ex}
\begin{tabular}{||c||c|c|c|c|c|c|c|}\hline
    {}  & ${a}^{(+)} $ & $C^a$ & $\mathcal{P}_a$ & $\phi_k^{(\mu)_{s-k}}$ & $C^{(\mu)_{s-1}}$  &
    $\phi^*_{k(\mu)_{s-k}}$ &$C^*_{(\mu)_{s-1}}$ \\ \hline
         $gh_{H}$ & 0 & 1 & -1  & 0  & 0 & 0& 0 \\
    \hline $gh_{L}$  & 0 & 0 & 0 & $0$ & $1$ & $-1$ & $-2$ \\
    \hline $gh_{\mathrm{tot}}$  & 0 & 1 & -1 & $0 $& $1$ &$-1$ &$-2$ \\
   \hline  $\epsilon$ & $0$
    & $1$ & $1$ & $0$&
    $1$  &$1$&$0$\\
   \hline
   \end{tabular}
\end{center} \vspace{-1ex}\caption{Ghost numbers and Grassman parity distributions.\label{table gh} }\end{table}
The \emph{classical antifield}  $|\chi^{0\ast}_{c}\rangle_s$ and
ghost field vectors $|C\rangle_{s}$   are naturally constructed
from the \emph{classical field} vector
 $|\chi^0_{c}\rangle_s$ and the gauge parameter
 $|\chi^1_{c}\rangle_s$, with the spin and ghost number
 relations  preserved on a basis of the correspondence \begin{eqnarray}\label{antistr}
   && \eta_0^{n_{f 0}}\eta_1^{+n_{f }}\mathcal{P}_1^{+n_{p }}|\phi(a^+)^{n_{f 0} n_{f }n_{p }}_{c}\rangle \to   \eta_0^{(n_{f 0}+1){} \mathrm{mod}{} 2}\mathcal{P}_1^{+n_{f }}\eta_1^{+n_{p }}|\phi^* (a^+)^{n_{f 0} n_{f }n_{p }}_{c}\rangle, \\
   \label{Cgaugep}
   && |\chi^1_{c}\rangle_s\  = \  |C\rangle_{s} \mu \ \ \texttt{for} \ \  (gh_H, gh_L, gh_{tot})\mu \ = \ (0,-1,-1).
 \end{eqnarray}
The minimal BV actions with account of  the constrained and complex conjugation properties  for the ghost field $C^{(\mu)_{s-1}}$ and antifields,
   \begin{eqnarray}
   && l_{11}\left(|\phi^*\rangle; |\phi^*_1\rangle, |\phi_2^*\rangle , |C^{(*)}\rangle\right) = (|\phi^*_2\rangle; 0,0,0 ), \label{constraintstar}\\
   && \label{complexcon}
     \big(C^{(\mu)_{s-1}}\big)^+ = C^{(\mu)_{s-1}},\ \  \big(\phi^{\ast}_{k(\mu)_{s-k}}\big)^+ = (-1)^k\phi^{\ast}_{k(\mu)_{s-k}}, \ k=0,1,2
   \end{eqnarray}
   in the general form, as well as the $\eta_0$-independent,
component, tensor-triplet $S_{C|s}$, doublet ${S}^d_{C|s}$, and
Fronsdal-like ${S}_{F|s}$ forms read as follows:
\begin{eqnarray}
 &&  S_{C|s}  =  \int d \eta_0 \; {}_{s}\langle \chi^0_{\mathrm{g}|c}
| Q_{c}| \chi^0_{\mathrm{g}|c} \rangle_{s} \  = \  {\cal{}S}_{C|s}  +  \int d \eta_0 \;\Bigl\{{}_{s}\langle \chi^{0\ast}_{c}|Q_c
| C \rangle_{s}+{}_{s}\langle C| Q_c
| \chi^{0\ast}_{c} \rangle_{s} \Bigr\} \label{Sgenfin1}\\
&&  \phantom{S_{C|s}}=
\mathcal{S}_{C|s} +
  \Biggr({}_{s}\left(\langle S^{*}_c \big|  {} \langle B^{*}_c \big|  \right)\left(\begin{array}{cc}
  l_0 & - \Delta Q_c\\
 -\Delta Q_c & \eta_1^+\eta_1            \end{array}\right)
       \left(  \begin{array}{c}\big|C\rangle_s \\  0   \end{array} \right)
 + h.c.\Biggr) \label{Sgenfin2}\\
 && \phantom{S_{C|s}}  = \mathcal{S}_{C|s} +  \biggl(\left[{}_{s-1}\langle \phi^{*}_1 \big|l_0  - {}_{s}\langle \phi^{*}_0 \big|l_1^+ - {}_{s-2}\langle \phi^{*}_2 \big|l_1  \right] C_{s-1}(a^+)|0\rangle +  h.c.\biggr)\label{Sgenfin3}\\
 && \phantom{S_{C|s}}  = \mathcal{S}_{C|s} +   2 \frac{(-1)^s}{s!} \int d^dx s
 \biggl(\phi^*_{(\mu)_{s}}\partial^{\mu_s}+ \phi^*_{1(\mu)_{s-1}}\partial^2-\phi^*_{(\mu)_{s-2}} \partial_{\mu_{s-1}}\biggr)C^{(\mu)_{s-1}};\label{Sgenfin4}\\
 && {S}^d_{C|s} = \mathcal{S}^d_{C|s} -  \biggl(\left[   {}_{s}\langle \phi^{*}_0 \big|l_1^+ +  {}_{s-2}\langle \phi^{*}_2 \big|l_1  \right] C_{s-1}(a^+)|0\rangle +  h.c.\biggr);\label{Sgenfindup}\\
 &&  {S}_{F|s}(\phi, \phi^*,C)  = \mathcal{S}_{F|s}(\phi) +2 \frac{(-1)^s}{s!} \int d^dx \phi^*_{(\mu)_{s}}\partial^{\{\mu_s}C^{(\mu\})_{s-1}}  \label{Sgenfin5}.
\end{eqnarray}
The functional $S_{C|s}$   is invariant  with respect to the minimal Lagrangian BRST-like transformations (with a Grassmann-odd constant parameter $\mu$) for the field vectors $|\chi^0_c(x) \rangle, |C(x) \rangle$
\begin{eqnarray}
  \delta_\mu|\chi^0_c(x) \rangle_{s}
& =  & \mu \frac{\overrightarrow{\delta}}{\delta {}_{s}\langle \chi^*_c(x) \big|}S_{C|s}  \ = \  \mu Q_c | C(x) \rangle_{s} ,  \label{dxbrst0}  \\
 \delta_\mu|C(x) \rangle_{s}
& =  & \mu \frac{\overrightarrow{\delta}}{\delta {}_{s}\langle C^*(x) \big|}S_{C|s}  \ = \  0, \ \ \  \epsilon\left(\frac{\overrightarrow{\delta}}{\delta {}_{s}\langle\chi^*_c(x) \big|}, \frac{\overrightarrow{\delta}}{\delta {}_{s}\langle C^*(x) \big|}\right) = 1,
\label{dxbrst1}
\end{eqnarray}
with constant antifields $ \delta_B| \chi^{0\ast}_{c} \rangle =0$ (as well as for the duals $\langle \chi^0_c(x)|, \langle C(x)| $)   or, equivalently, in terms of a \emph{new BRST-like generator} $\overrightarrow{s}_0$ and its dual $\overleftarrow{s}_0$:
 \begin{eqnarray}\label{brsnewgen}
  && \delta_B \left[ |\chi^0_c(x) \rangle_{s} , |C(x) \rangle_{s}\right]\ =\  \mu \overrightarrow{s}_0\left[|\chi^0_c(x) \rangle_{s} , |C(x) \rangle_{s}\right]  \ =\ \mu\left[ Q_c  , 0\right] | C(x) \rangle_{s}, \\
    && \delta_B \left[ {}_{s}\langle\chi^0_c(x)|  , {}_{s}\langle C(x) | \right]\ =\   \left[{}_{s}\langle\chi^0_c(x)|  , {}_{s}\langle C(x) | \right] \overleftarrow{s}_0 \mu\ =\  {}_{s}\langle C(x) |\left[ Q_c  , 0\right]\mu . \label{brsnewgend}
\end{eqnarray}
Indeed,
\begin{equation}\label{BRSTinv}
  \delta_B S_{C|s} =  \left(\delta_B  {}_{s}\langle\chi^0_c| \frac{\overrightarrow{\delta}S_{C|s}}{\delta {}_{s}\langle\chi^0_c \big|}+ \frac{S_{C|s}\overleftarrow{\delta}}{\delta |\chi^0_c \rangle_{s}\big|}\delta_B |\chi^0_c\rangle_{s} \right) =\mu \Big({}_{s}\langle C | Q_c^2  |\chi^0_c\rangle_{s}-{}_{s}\langle \chi^0_c | Q_c^2  |C\rangle_{s} \Big)=0.
\end{equation}
The variational derivatives   with respect to the vectors $ |\chi^{0(*)}_c \rangle_{s}$, $|C^{(*)}(x) \rangle_{s}$ and their duals in (\ref{dxbrst0}),  (\ref{dxbrst1}), (\ref{BRSTinv}), e.g.  for any quadratic (in the fields) functional with the kernel $E_{F}$
  \begin{equation}\label{represfunc}
  F  = \int d \eta_0 \mathcal{F}(\chi^{0(*)}_c, C^{(*)}) = \int d \eta_0 {}_{s}\langle \chi^0_{\mathrm{g}|c}
| E_{F}| \chi^0_{\mathrm{g}|c} \rangle_{s} \equiv \int d \eta_0 \mathcal{F}(\eta_0)
\end{equation}
are given in terms of  variational derivatives for a fixed  $\eta_0$ with a vanishing Grassman parity of the density $\mathcal{F}$ ($\epsilon(\mathcal{F}) = \epsilon(E_{F})=\epsilon(F)+1$)  according to (\ref{genfstr}), (\ref{genfstr1})
\begin{eqnarray}
&& \hspace{-1em}\left(\frac{F \overleftarrow{\delta}}{\delta  |\chi^{0(*)}_{c} \rangle_s}; \frac{  \overrightarrow{\delta}F}{\delta {}_s\langle\chi^{(*)}_c |}; \frac{F \overleftarrow{\delta}}{\delta  |C^{*} \rangle_s}; \frac{  \overrightarrow{\delta}F}{\delta {}_s\langle C |}\right)   =   \left(\frac{{\mathcal{F}}\overleftarrow{\delta}_{\eta_0}}{\delta  |\chi^{0(*)}_{c} \rangle_s};
\frac{  \overrightarrow{\delta}_{\eta_0}{\mathcal{F}}}{\delta {}_s\langle\chi^{0(*)}_{c} |}; \frac{\mathcal{F}\overleftarrow{\delta}_{\eta_0}}{\delta   |C^{*} \rangle _s};
 \,\frac{  \overrightarrow{\delta}_{\eta_0}\mathcal{F}}{\delta {}_s\langle C  |}\right) . \label{transf1}
\end{eqnarray}
For these variational derivatives, the following normalization
 holds true (for $\delta(\eta'_0-\eta_0) = \eta'_0-\eta_0$):
  \begin{align}
& \left( \frac{|A(\eta_0;x) \rangle_s  \overleftarrow{\delta}}{\delta  |A(\eta'_0;x') \rangle_s}; \frac{  \overrightarrow{\delta}{}_s\langle A (\eta_0;x)\big|}{\delta {}_s\langle A (\eta'_0;x')\big|}\right)  =  \delta(\eta'_0-\eta_0) \big(\delta(x'-x);  \delta(x'-x)\big), \  A\in \{{\chi}^{0(*)}_{c}, {C}^{(*)}_{c}\}.\label{supus1}\end{align}
The BRST-BV  actions allow for consistency when
deriving interaction vertexes (for developments in the metric-like form see, e.g.,  \cite{reviews3}, \cite{BRST-BV3}, \cite{BFPT}, \cite{DT}). In
general (e.g. for the cubic vertex), one considers three
independent Hilbert spaces, $\mathcal{H}{}^i_{c|tot} =
\mathcal{H}_i\otimes H^{o_a}_{i|gh}$, $i=1,2,3$, and finds a BRST
invariant vertex $V$ in the tensor product
$\otimes_{i=1}^3\mathcal{H}{}^i_{c|tot}$ without any off-shell
constraints \cite{BFPT}, being the case of a reducible
$ISO(1,d-1)$ representation. Restricting ourselves for simplicity
of illustration by a cubic self-interacting  vertex,
$\big|{V}\rangle_{(s,s,s)} = $
$\big|{V}(\{\widetilde{l}{}_0\},\{\widetilde{l}{}_1^+\},\{\widetilde{l}{}^+_{11}\})
\rangle_{(s,s,s)}\equiv $ $\big|{V}  \rangle_{3s}$, we can solve
the problem, for instance, by deforming $\mathcal{S}_{F|s}(\phi)$
(\ref{SclsrsingleF}) and the gauge transformations
(\ref{gaugetrsing}) within a gauge model having the same
double-traceless field $\phi_{(\mu)_s}$ by means of the
self-interaction terms $\mathcal{S}_{1|s}(\phi)$ and by $S_{1g|s}
= S_{1g|s}(\phi, \phi^*, C)$, respectively,
\begin{eqnarray}\label{BVFronscacint}
  \hspace{-1em} &\hspace{-1em}&\hspace{-1em}{S}_{[1]|s}(\phi_{}, \phi^*, C,C^*)   =  {S}_{F|s}(\phi, \phi^*, C) +  \mathcal{S}_{1|s}(\phi)+  S_{1g|s} + S_{2g|s}(\phi, \phi^*, C,C^*), \\
    \hspace{-1em} &\hspace{-1em}&\hspace{-1em}\mathcal{S}_{1|s}(\phi) =  g\Big({}_{s}\langle{\phi}\big|  \otimes {}_{s}\langle{\phi}\big|\otimes {}_{s}\langle{\phi}\big|\big|{V}  \rangle_{3s} +  {}_{3s}\langle {V}^+\big|  \big|\phi \rangle_{s}\otimes \big|\phi \rangle_{s}\otimes \big|\phi \rangle_{s}\Big) ,  \label{Fronscacint}\\
     \hspace{-1em} &\hspace{-1em}&\hspace{-1em} S_{1g|s} =  g\Big({}_{s}\langle{\phi^*}\big| \otimes  {}_{s}\langle{\phi}\big|{}_{s-1}\langle{C}\big| \big|{V_1}\big(\{\widetilde{l}_0\}\{\widetilde{l}_1\},\{\widetilde{l}_{11}\}\big)  \rangle_{3s-1} + {}_{3s-1}\langle {V_1^+} \big| \big|C \rangle_{s-1} \otimes \big|\phi \rangle_{s} \otimes \big|\phi^* \rangle_{s}\Big),  \label{gaugFronsint}\\
      \hspace{-1.0em} &\hspace{-1.0em}&\hspace{-1.0em} S_{2g|s}= \frac{g}{2}\Big( {}_{s-1}\langle{C^*}\big| \otimes {}_{s-1}\langle{C}\big| \otimes{}_{s-1}\langle{C}\big|  \big|  {F} \rangle_{3(s-1)} +   {}_{3(s-1)}\langle{F}^+ \big | \big |  C \rangle_{s-1}\otimes \big | C \rangle_{s-1}\otimes \big | C^* \rangle_{s-1} \Big).\label{gaugalgFrons}
\end{eqnarray}
Here, first,    the local product $\otimes_{k=1}^p |\phi \rangle_{s}$ (and also  the sets $\{\widetilde{l}{}_0\}$, $\{\widetilde{l}_{1}\}$,  $ \{\widetilde{l}_{11}\}$) is understood as
\begin{eqnarray*}
&&\otimes_{k=1}^p \big(\phi_{(\mu^k)_s}(x)\prod a^{+{(\mu^k)_s}}_k|0\rangle\big)\ \ \  \mathrm{ and} \ \ \  \{\widetilde{l}{}_0\} = \{\eta^{\mu\nu}\partial^k_\nu\partial^k_\mu\} \equiv \{l_0^k\}, \\
&& \{\widetilde{l}_{1}\} =\{ -\imath a^{\mu,k}\partial^l_\mu\}\equiv \{l_1^{kl}\}, \qquad   \{\widetilde{l}_{11}\}= \{\textstyle\frac{1}{2} a^{\mu,k}a^l_{\mu}\}\equiv \{l_{11}^{kl}\} , \ k, l =1,...,3,\end{eqnarray*}
with 3 sets of oscillators $a^{+\mu}_k, a^{\nu}_k$, $[a^{+\mu}_k, a^{\nu}_l]=\delta_{kl}\eta^{\mu\nu}$. Second,  the  last summand  (\ref{gaugalgFrons}) with an operator quantity $\big|  {F} \rangle_{3(s-1)} \equiv \big|  {F}\big(\{\widetilde{l}_0\},\{\widetilde{l}_1\},\{\widetilde{l}_{11}\}\big) \rangle_{3(s-1)} $ is  necessary to have the closed
deformed algebra of  non-Abelian gauge transformations   determined with  help of field independent term, $\big|{V_1}\big(\{\widetilde{l}_0\},\{\widetilde{l}_1\},\{\widetilde{l}_{11}\}\big)\rangle_{3s-1} $:
\begin{eqnarray}\label{defgtrFr}
  &&\delta_{[1]} |\phi \rangle_{s} =  (\delta_{0}+\delta_{1}  )|\phi \rangle_{s} = l_1^+|\xi\rangle_{s-1} + g\Big({}_{s}\langle{\phi}|\otimes{}_{s-1}\langle{\xi}\big| +{}_{s-1}\langle{\xi}\big|\otimes{}_{s}\langle{\phi}|\Big)\big|{V_1}  \rangle_{3s-1}, \\
 &&  \big|{V_1}  \rangle_{3s-1} =   \sum_{k,l,m,i,j=1}^{3} \sum_{n_k, n_{ij},n_{lm}} V_{1|n_{k}n_{ij}n_{lm}}\big(l_0^k\big)^{n_{k}}\big(l_1^{+ij}\big)^{n_{ij}}\big(l_{11}^{+lm}\big)^{n_{lm}}|0\rangle\otimes|0\rangle\otimes|0\rangle
\label{V0}\end{eqnarray}
  with  dimensionless coupling constant $g$,  with integers $n_{k}, n_{ij}, n_{lm}$ satisfying to the requirement of dimensionless of the action and to spin the condition: $n_{ij}+ 2n_{lm}=3s-1$.  We imply the same representations with unknown  coefficients $V_{n_{k}n_{ij}n_{lm}}$, $F_{n_{k}n_{ij}n_{lm}}$
with respective restrictions on the integers for the vertex $|V\rangle_{3s}$ and for the structure constant $|F\rangle_{3(s-1)}$.

A  consistent deformation of the free BRST-BFV action $\mathcal{S}_{F|s}(\phi)$ leads to the relations
  \begin{equation}\label{consistency relations}
    \delta_{1}\mathcal{S}_{F|s}(\phi) +\delta_{0} \mathcal{S}_{1|s}(\phi)   =  0 ,   \quad  \big[\delta_{[1],\xi_1},\, \delta_{[1],\xi_2}\big]  |\phi \rangle_{s} = \delta_{[1],\xi_3}|\phi \rangle_{s}  + {o}(\phi^2), \ \xi_3=\xi_3(\xi_1,\xi_2).
  \end{equation}
Equivalently, the consistency in the deformation of the classical
action is to be controlled by the solvability of the master
equation for a deformed BRST--BV action  in the language
of a component antibracket $(\bullet,\bullet)$ = $\frac{\overleftarrow{\delta} \bullet}{\delta \phi^A_{\min}}\frac{\overrightarrow{\delta} \bullet}{\delta \phi^*_{A\min}} - \frac{\overleftarrow{\delta} \bullet}{\delta \phi^*_{A\min}}\frac{\overrightarrow{\delta} \bullet}{\delta \phi^A_{\min}}$,
\begin{equation}\label{antibracketdef}
   \left({S}_{[1]|s}, {S}_{[1]|s}\right)  = 2 \int d^d x \bigg(\frac{\overleftarrow{\delta} {S}_{[1]|s}}{\delta \phi^{(\mu)_s}(x)}\frac{\overrightarrow{\delta} {S}_{[1]|s}}{\delta \phi^*_{(\mu)_s}(x)}+ \frac{\overleftarrow{\delta} {S}_{[1]|s}}{\delta C^{(\mu)_{s-1}}(x)}\frac{\overrightarrow{\delta} {S}_{[1]|s}}{\delta C^*_{(\mu)_{s-1}}(x)} \bigg) = 0.
\end{equation}
A detailed consideration of self-interaction and interaction
vertices (involving TS HS fields of different helicities\footnote{See, recently considered  in \cite{2011.02689}  the deformation of the minimal BV actions in the single-field (Fronsdal) formulation, first being the sum for two free double traceless HS fields of helicities $s, 2s$, second the sum of three HS fields of helicities  $s_1, s_2, s_1+s_2$ on a base of resolution of the master equation (\ref{antibracketdef}) for deformed BV action with cubic vertexes.})
according to the proposed algorithm poses a separate problem.

Let us now turn ourselves to a non-minimal extension of the
BRST--BV approach for the TS HS field in question.

\section{Non-minimal BRST-BV Lagrangians}
\label{nonBRSTBV} 

Since all the monomials amongst the minimal BFV ghost oscillators
have been already utilized to compose a minimal generalized vector
$|\chi^0_{\mathrm{g}|c}\rangle$ (\ref{genfstr}), we need to
enlarge the concept of BFV--BV duality beyond the minimal sector.
In the first place, we augment our constrained BRST--BV approach
by a Lagrangian $S_{C|s}$  (\ref{Sgenfin1}) for a field of spin
$s$ in the triplet form, by introducing BFV non-minimal
oscillators of antighosts $\overline{C}{}^a$,
$\overline{\mathcal{P}}_a$ and Lagrangian multipliers $\lambda^a$,
$\pi_a$ according to the numbers $N_{\mathrm{nmin}}=4(n_{o_a}-1)
=8$ (with no allowance for $l_0$) in order to present a total
constrained BRST operator $Q_{c|tot}$ with the properties (for a
vanishing $gh_L$)
\begin{eqnarray}\label{nmingras}
&& \begin{array}{||c||c|c|c|c|c|c|c|c|}\hline
     & \overline{\eta}{}_1 & \overline{\eta}{}^+_1 & \overline{ \mathcal{P}}{}^+_1 & \overline{ \mathcal{P}}{}_1 & \lambda_1 & \lambda^+_1 & \pi_1 & \pi_1^+ \\ \hline
     \epsilon & 1 & 1 & 1 & 1 & 0 & 0 & 0 & 0 \\ \hline
     gh_H & -1 & -1 & 1 & 1 & 0 & 0 & 0 & 0 \\
\hline
   \end{array} ,\\
  &&   \{\overline{\eta}{}_1, \,\overline{\mathcal{P}}{}^+_1\big\}=\{\overline{\eta}{}^+_1, \,\overline{\mathcal{P}}{}_1\big\}= 1,\   \big[\lambda_1, \,\pi_1^+\big] = \big[\pi_1, \,\lambda_1^+\big] = 1. \label{nmincommrel}
\end{eqnarray}
The operator $Q_{c|tot}$ depending on the whole set of BFV
oscillators $\Gamma_{gh}$ required to construct a unitarizing
Hamiltonian for a topological  dynamical  system \cite{BFV}, as
well as a total BRST-extended constraint $\mathcal{L}_{11}$ and a
spin operator  $\sigma_{c|tot}$, which act on a total Hilbert
space $\mathcal{H}^{\mathrm{nmin}}_{g|\mathrm{tot}}\equiv
\mathcal{H}_{g|\mathrm{tot}}\otimes \mathcal{H}_{\mathrm{nmin}}$
are  found  as solutions of \emph{non-minimal generating
equations} of the form (\ref{geneq}):
\begin{eqnarray}\label{Qctot}
&& Q_{c|tot} \ =\  Q_{c} + \overline{ \mathcal{P}}_1\pi^+_1+ \pi_1\overline{ \mathcal{P}}{}^+_1 , \\
 && \mathcal{L}_{11} \ = \ \widehat{L}_{11}+ \overline{\eta}_{1} \overline{\mathcal{P}}_{1}+ \lambda_1\pi_1, \ \ \sigma_{c|tot}\ = \ \widehat{\sigma}_c +\overline{\eta}{}^+_{1} \overline{\mathcal{P}}_{1}- \overline{\eta}_{1}\overline{\mathcal{P}}{}^+_{1} + \lambda_1^+\pi_1+ \lambda_1\pi^+_1. \label{Lsigmactot}
\end{eqnarray}
Second, we extend the  minimal BRST-BV approach to a non-minimal
one by introducing a  tensor  antighost, a
Nakanishi-Lautrup fields and their antifields of the non-minimal
BV sector (for a triplet LF), with the respective $(\epsilon,
gh_L)$ distributions (for $gh_H \equiv  0$)
\begin{eqnarray}\label{confspnm}
 &&  \{\phi^A\}  = \big\{ \phi^A_{\min};  \overline{C}{}^{(\mu)_{s-1}}(x),\, B{}^{(\mu)_{s-1}}(x)\big\} , \ \  \{\phi^*_{A}\} = \big\{ \phi^*_{A\min};  \overline{C}{}^*_{(\mu)_{s-1}}(x), B{}^*_{(\mu)_{s-1}}(x)\big\},\\
 \label{nminghL}
&& \qquad \begin{array}{||c||c|c|c|c|}\hline
     & \overline{C}{}^{(\mu)_{s-1}} & B{}^{(\mu)_{s-1}} & \overline{C}{}^*_{(\mu)_{s-1}} &B{}^*_{(\mu)_{s-1}} \\ \hline
     \epsilon & 1 & 0 & 0 & 1 \\ \hline
     gh_L & -1 & 0 & 0 & -1  \\
\hline
   \end{array}.
\end{eqnarray}
These (anti)fields  are multiplied  (inside the respective
monomials of a \emph{total generalized vector}
$|\chi^{0}_{\mathrm{tot}|c}\rangle_s \in
\mathcal{H}^{0|\mathrm{nmin}}_{g|\mathrm{tot}}$) only by
non-minimal BFV oscillators (where
$\mathcal{H}^{0|\mathrm{nmin}}_{g|\mathrm{tot}}\hspace{-0.1em}  $
$\equiv$
$\hspace{-0.1em}\sum_{e+l=0}\mathcal{H}^{e,l|\mathrm{nmin}}_{g|\mathrm{tot}}$
for  $\mathcal{H}^{\mathrm{nmin}}_{g|\mathrm{tot}}$ admitting the
natural $\mathbb{Z}\oplus \mathbb{Z}$ grading
$\mathcal{H}^{\mathrm{nmin}}_{g|\mathrm{tot}}= \sum_{e,l\geq
0}\mathcal{H}^{e,l|\mathrm{nmin}}_{g|\mathrm{tot}}$), namely,
\begin{eqnarray}
\hspace{-1em} &\hspace{-1em}&\hspace{-1em} |\chi_{\mathrm{tot}|c} \rangle =  \sum_{n}\frac{\imath^{n}}{ n!} \prod ( \eta_0
)^{n_{f 0}}  \prod( \eta_1^+ )^{n_{f }} (
\mathcal{P}_1^+ )^{n_{p}}( \overline{\eta}_1^+ )^{n_{\bar{f}}} (
\overline{\mathcal{P}}_1^+ )^{n_{\bar{p} }} \prod( \lambda_1^+ )^{n_{\lambda }} (
\pi_1^+ )^{n_{\pi }} \phi
^{N_{\mathrm{tot}}}_{\mathrm{tot}|c} ( a^+) |0\rangle. \label{chirtot}
\end{eqnarray}
with a chosen representation $(\overline{\eta}_1, \overline{\mathcal{P}}_1, \lambda_1, \pi_1)|0\rangle= 0$  and $N_{\mathrm{tot}} \equiv  (n_{f 0}, n_{f },n_{p },n_{\bar{f} }, n_{\bar{p} }, n_{\lambda },n_{\pi })$.

Once again, the  same spectral problem (\ref{BRSTcomplex}), albeit
for the $Q_{c|\mathrm{tot}}$-complex with imposed spin, the
BRST-extended constraint (\ref{Lsigmactot}) in
$\mathcal{H}^{\mathrm{nmin}}_{g|\mathrm{tot}}$ leads to the
representation
\begin{eqnarray}
\hspace{-1em} &\hspace{-1em}&\hspace{-1em} |\chi^{{}0}_{\mathrm{tot}|c}\rangle_s  =   |\chi^0_{\mathrm{g}|c}\rangle_s + \overline{\mathcal{P}}_1^+ \big|\overline{C}(a^+)\rangle_{s-1}+ \lambda^+_1 \big|b(a^+)\rangle_{s-1}+ \eta_0\big(\overline{\eta}_1^+ \big|\overline{C}{}^*(a^+)\rangle_{s-1}+  \pi^+_1 \big|b^*(a^+)\rangle_{s-1}\big)  \nonumber\\
\hspace{-1em} &\hspace{-1em}&\hspace{-1em} \phantom{|\chi^{{}0}_{\mathrm{tot}|c}\rangle_s} \equiv   |\chi^0_{\mathrm{g|c}}\rangle_s +  \big|\overline{C}(\overline{\mathcal{P}}_1^+,a^+)\rangle_{s}+  \big|b(\lambda^+_1,a^+)\rangle_{s}+ \big|\overline{C}{}^*(\overline{\eta}_1^+,a^+)\rangle_{s}+   \big|b^*(\pi^+_1,a^+)\rangle_{s}\label{gentot0}
\label{Stot}\end{eqnarray}
[$(\epsilon,  gh_{tot}) |\chi^{0}_{\mathrm{tot}|c}\rangle = (0,0)$].   Here, we have used the rule (\ref{antistr}) augmented  in the  non-minimal sector to construct  antifield vectors for $ \big|\overline{C}(\overline{\mathcal{P}}_1^+,a^+)\rangle$ and $\big|b(\lambda^+_1,a^+)\rangle_{s}$.  As a result, the Grassmann-even functional
\begin{eqnarray}
&&  {S_{0|s}(\chi^{0|\mathrm{nmin}}_{\mathrm{tot}|c}) \ = \ \int d \eta_0 \; {}_{s}\langle \chi^{0|\mathrm{nmin}}_{\mathrm{tot}|c}
\big| Q_{c|\mathrm{tot}}\big| \chi^{0|\mathrm{nmin}}_{\mathrm{tot}|c}\rangle_{s}
, \quad  \mathcal{L}_{11}\big|\chi^{0|\mathrm{nmin}}_{\mathrm{tot}|c}\rangle_s=0}
\end{eqnarray}
  is nothing else  than  the BV action $S_{0|s}= S_{ext}(\phi, \phi^*)$ in the constrained formulation with allowance for the fact that $(\mathcal{L}_{11}, \sigma_{c|tot}) |\chi_{\mathrm{g}|c}\rangle$  =  $(\widehat{L}_{11}, \widehat{\sigma}_c) |\chi_{\mathrm{g}|c}\rangle$.  The action $S_{0|s}$ in the
   $\eta_0$-independent,  ghost-independent, tensor triplet   representations  acquires the forms
\begin{eqnarray}
 &&   S_{0|s}   =    {S}_{C|s}  +  \int d \eta_0 \;\Bigl\{{}_{s}\langle \overline{C}{}^*(\overline{\eta}_1^+,a^+)\big|\big(\overline{ \mathcal{P}}_1\pi^+_1+ \pi_1\overline{ \mathcal{P}}{}^+_1\big)
| b(\lambda^+_1,a^+) \rangle_{s}+h.c. \Bigr\} \label{SBVfin1}\\
&&  \phantom{S_{C|s}}=
{S}_{C|s} - {}_{s-1}\langle \overline{C}{}^*(a^+)\big| b(a^+) \rangle_{s-1}-{}_{s-1}\langle b(a^+)\big| \overline{C}{}^*(,a^+) \rangle_{s-1}
 \label{SBVfin2}\\
 && \phantom{S_{C|s}}  =  {S}_{C|s} +   2 \frac{(-1)^s}{s!} \int d^dx \ s \overline{C}{}^*_{(\mu)_{s-1}}b^{(\mu)_{s-1}}\label{SBVfin4}
\end{eqnarray}
 with traceless tensor  (anti)fields from non-minimal sector (for $\mathbb{R}$-valued $\overline{C}{}^*_{(\mu)_{s-1}}$, $b^{(\mu)_{s-1}}$ and pure imaginary $\big(\overline{C}^{(\mu)_{s-1}}$, $b{}^*_{(\mu)_{s-1}}\big)^+$ = $- \big(\overline{C}^{(\mu)_{s-1}}$, $b{}^*_{(\mu)_{s-1}}\big)$).

Notice that the doublet and single-field forms of the BV actions
 are determined by the functionals ${S}^d_{C|s}$ (\ref{Sgenfindup}),
 ${S}_{F|s}$ (\ref{Sgenfin5}), with the same non-minimal extension
 $\overline{C}{}^*b$. The functional $S_{0|s}$   satisfies
 the master equation (\ref{antibracketdef}) with
 an  appropriate antibracket written for the respective
 (triplet, doublet or single-field) representation
 in the total field-antifield space.
 It  is invariant with respect to the  Lagrangian BRST
 transformations of the field vector
 $|\chi^{{}0}_{\mathrm{f}|c}\rangle$,
 when presenting  $ |\chi^{{}0}_{\mathrm{tot}|c}\rangle
 =  |\chi^{{}0}_{\mathrm{f}|c}\rangle+
 |\chi^{{}0}_{\mathrm{af}|c}\rangle$
 for $|\chi^{{}0}_{\mathrm{(a)f}|c}\rangle$ depending
 on (anti)fields for ket-vector and its dual:
\begin{eqnarray}
 \hspace{-0.5ex} \delta_B |\chi^0_{\mathrm{f}|c}(x) \rangle_{s}
\hspace{-0.5ex}& \hspace{-1ex}=  & \hspace{-1ex} \mu \frac{\overrightarrow{\delta}}{\delta {}_{s}\langle \chi^0_{\mathrm{af}|c}(x) \big|}S_{0|s}  \ = \  \mu\Big(Q_{c} | C(x) \rangle_{s} + \big[\overline{ \mathcal{P}}_1\pi^+_1+ \pi_1\overline{ \mathcal{P}}{}^+_1\big]\big|b(\lambda^+_1,a^+,x)\rangle_{s} \Big) ,  \label{totbrst0}  \\
\hspace{-0.5ex}\delta_B S_{0|s} \hspace{-0.5ex}& \hspace{-1ex}=  & \hspace{-1ex}   \Big(\delta_B  {}_{s}\langle\chi^0_{\mathrm{f}|c}| \frac{\overrightarrow{\delta}S_{0|s}}{\delta {}_{s}\langle\chi^0_{\mathrm{f}|c} \big|}+ \frac{S_{0|s}\overleftarrow{\delta}}{\delta |\chi^0_{\mathrm{f}|c} \rangle_{s}\big|}\delta_B |\chi^0_{\mathrm{f}|c}\rangle_{s} \Big)  = \delta_B S_{C|s} +  \delta_B\big(S_{0|s}-S_{C|s}\big)=0. \label{totbrstact}
\end{eqnarray}
In  a ghost-independent  form,  the BRST transformations take the
usual form, as we omit the symbol $"s"$
\begin{equation}\label{compBRST}
   \delta_B \Big[\left( \big|\phi\rangle ,      \big|\phi_1\rangle,    \big|\phi_2\rangle\right),  \big|{C}(a^+)\rangle,  \big|\overline{C}(a^+)\rangle,  \big|b(a^+)\rangle\Big] = \Big[\left( l_1^+ ,l_0,l_1\right) \big|C(a^+)\rangle, 0, \big|b(a^+)\rangle, 0\Big] \mu.
\end{equation}
\section{Gauge-fixing and BRST-invariant  quantum action}

\label{fgBRSTBV} 

To determine a non-renormalized quantum action
$S_{0}^\Psi(\chi^{\mathrm{tot}{}0}_{g|c}) $, we  introduce a
quadratic  gauge-fermion  functional
$\Psi(\chi^{{}0}_{\mathrm{tot}|c})$  corresponding to
$R_{\xi,\beta}$-gauges with the help of Grassman-even $x$-local
kernel $\widehat{E}^{\Psi}_{\xi,\beta}$  constructed from BRST-BFV
operator gauge-fermion  $\Psi_H(\overline{\eta}^{(+)}_1,
\pi^{(+)}_1,o_a; \xi, \beta) $:
\begin{eqnarray}
\hspace{-1em} &\hspace{-1em}&\hspace{-1em} \Psi\left(\chi^{{}0}_{\mathrm{tot}|c}  \right)  \ = \            \int d \eta_0 \; {}_{s}\langle \chi^{0}_{\mathrm{tot}|c}
\big| \widehat{E}^{\Psi}_{\xi,\beta}\big| \chi^{0}_{\mathrm{tot}|c} \rangle_{s}, \     \ \mathrm{for}\ \ \widehat{E}^{\Psi}_{\xi,\beta} \stackrel{def}{=} \eta_0 \Psi_H {=}  \eta_0  \Psi_H(o_a, \Gamma_{gh};\xi, \beta), \label{solmastereq3}\\
 \hspace{-1em} &\hspace{-1em}&\hspace{-1em} \Psi_H \ = \    \overline{\eta}^+_1\left(l_1+ \mathcal{P}_1\eta_1\Big[(1+\beta)l_1^+ + \frac{2\beta}{2s-4+ d} l_1l^+_{11}\Big]+
 \frac{\xi}{2}\pi_1\right) - h.c. \equiv \Psi_H^0 -\big(\Psi_H^0 \big)^+, \label{solmastereq3ferm}
\end{eqnarray}
with
$(\epsilon, gh_H, gh_L, gh_{\mathrm{tot}}) \Psi_H= (1,-1,0,-1)$
and $l^+_{11}=(1/2)a^+_{\mu}a^{+\mu}$.
The property of anti-hermitian conjugation    $\Psi_H^+ =-\Psi_H$
provides the hermiticity of  $\widehat{E}^{\Psi}_{\xi,\beta}$:
$\big(\widehat{E}^{\Psi}_{\xi,\beta}\big)^+=\widehat{E}^{\Psi}_{\xi,\beta}$.
The definition of  $\Psi_H$ as
$\Psi_H = \overline{C}^a \chi_a $  (\ref{solmastereq3ferm})
  determines the first-order operator $\widehat{\chi}_a \equiv \widehat{\chi}$
   and $\widehat{\chi}{}^+$ of the gauge condition
\begin{equation}\label{gaugegen}
\widehat{\chi}\big(\big| \chi^{0}_{c} \rangle_{s}+\big|b(\lambda^+_1)\rangle_{s}\big) = 0, \quad \big({}_{s}\langle \chi^{0}_{c} \big|+{}_{s}\langle b(\lambda_1)\big|\big)\widehat{\chi}{}^+=0
\end{equation}
 being equal in number with that of the independent gauge parameters $ |\chi^1_c\rangle_s$ (\ref{parctotsym}).
 The   gauge condition (\ref{gaugegen}), as expanded
 in ghost powers, is equivalent to  three equations
 \begin{equation}\label{gaugecomp}
   l_1 |\phi\rangle_s - \Big[(1+\beta)l_1^+ + \frac{2\beta}{2s-4+ d} l_1l^+_{11}\Big]|\phi_2\rangle_{s-2}+
 \frac{\xi}{2}|b\rangle_{s-1}= 0; \  \ l_1|\phi_k\rangle_{s-k}=0, k=1,2.
 \end{equation}
where the equation for $|\phi_1\rangle_{s-1}$ does not
contribute to $\Psi(\chi^{{}0}_{\mathrm{tot}|c})$ due to
$\eta_0^2\equiv 0$. For $(\xi,\beta)=0$  and $(\xi,\beta)=(1,0)$,
the gauge (\ref{gaugegen})  corresponds to the Landau and Feynman
gauges (for $s=1$), respectively. Any of the gauge conditions in
(\ref{gaugecomp}) respects the property of tracelessness (which
means $l_{11}\big(|\phi\rangle_s - ...\big) =0$ and
$l_{11}l_1|\phi_k\rangle =0$ on the traceless constraint surface).
In the ghost dependent form it is equivalent to the relation (with
allowance for $[l_{11},\,l^+_{11}]|\phi_k\rangle_{s-k} =
\{g_0-(s-k+d/2)\}|\phi_k\rangle_{s-k}=0$)
 \begin{equation}\label{genferm}
  [\widehat{\chi},\, \mathcal{L}_{11}\}\big(\big| \chi^{0}_{c} \rangle_{s}+\big|b(\lambda^+_1)\rangle_{s}\big) =  \mathcal{P}_1\eta_1  l_1\big| \chi^{0}_{c} \rangle_{s}\stackrel{l_1|\phi_2\rangle =0}{=}0.
\end{equation}
In the ghost-independent representation, the gauge-fermion  functional
$\Psi(\chi^{{}0}_{\mathrm{tot}|c})$ reads (we omit the spin index)
\begin{eqnarray}
\hspace{-1em} &\hspace{-1em}&\hspace{-1em} \Psi\big(\phi_k, \overline{C}, b  \big)   \hspace{-0.1em}  =  \hspace{-0.1em}    \bigg\{\hspace{-0.1em} \langle \overline{C}(a)
\big| \Big(l_1\big|\phi\rangle -\big[(1+\beta) l_1^+ + \frac{2\beta}{2s-4+ d} l_1l^+_{11}\big]\big|\phi_2\rangle +   \frac{\xi}{2} \big|b(a^+)\rangle  \hspace{-0.1em} \Big)- h.c. \hspace{-0.1em} \bigg\}. \label{solmastereq5}
\end{eqnarray}
In the single-field (Fronsdal) formulation, the functional
(\ref{solmastereq5}) transforms as  $\Psi_{F|s}$ $\equiv$
$\Psi\big(\phi, \overline{C}, b  \big)$ = $\Psi\big(\phi_k,
\overline{C}, b  \big)\vert_{|\phi_2\rangle =
-l_{11}|\phi\rangle)}$  and reads in the tensor form as
\begin{eqnarray}
&&\Psi_{F|s}= - 2 \frac{(-1)^{s}}{s!} \hspace{-0.15em}\int \hspace{-0.15em}d^dx \ s \overline{C}{}^{(\mu)_{s-1}}\bigg\{ \partial^{\mu_s}\phi_{(\mu)_s} + (s-1)(1+\beta)\partial_{\mu_s-1}\phi_{(\mu)_{s-2}}{}^\nu{}_\nu \label{solmastereq55}\\
&& \phantom{\Psi_{F|s}}+ \frac{2\beta }{2s-4+ d}(s-1)\big[\partial_{\mu_s-1}\phi_{(\mu)_{s-2}}{}^\nu{}_\nu +  \frac{1}{2}(s-2) \eta_{\mu_{s-1}\mu_{s-2}}\partial^{\rho}\phi_{(\mu)_{s-3}\rho}{}^\nu{}_\nu \big]+  \frac{\xi}{2}b_{(\mu)_{s-1}} \bigg\}. \nonumber
\end{eqnarray}
The ghost-dependent Grassmann-odd density
${\mathcal{M}}_{\Psi|c}(x,y)\equiv$
${\mathcal{M}}_{\Psi|c}(\Gamma_{gh}, a^{(+)},x,y)$ of the
Fadde\-ev-Popov operator ${M}_{\Psi|c} (x,y)$, which  is implied by a
 variation of the gauge condition  (in terms of $ \Psi_H^0$)
under the gauge  transformation $\delta\big(\Psi_H^0\big|
\chi^{0}_{c}(x) \rangle \big)$  (in order to extract a single
representative from a gauge orbit) admits the representation
\begin{eqnarray}
\hspace{-1em} &\hspace{-1em}&\hspace{-1em}   {\mathcal{M}}_{\Psi|c}(x,y)
\stackrel{def}{=}   \eta_0 \hspace{-0.2em}\int \hspace{-0.2em}d^dy\bigg\{\hspace{-0.15em}\Psi_H^0\big| \chi^{0}_{c}(x) \rangle_{s}  \frac{\overleftarrow{\delta}_{\eta_0}}{\delta \big| \chi^0_c(y) \rangle_{s}}Q_c(y) -  Q_c(y)\frac{\overrightarrow{\delta}_{\eta_0}}{\delta{}_{s}\langle\chi^{0}_{c}(y)\big|}\Big({}_{s}\langle\chi^{0}_{c}(x)\big|  (\Psi_H^0)^+\Big)\hspace{-0.15em}\bigg\} \label{FPdens} \\
\hspace{-1em} &\hspace{-1em}&\hspace{-1em}   \phantom{{\mathcal{M}}_{\Psi|c}(x,y)\ }   = \eta_0\int \hspace{-0.2em}d^dy \Big( \Psi_H^0\vert_{\xi=0}(x) \delta(x-y)Q_c(y)-Q_c(y)(\Psi_H^0)^+\vert_{\xi=0}(x) \delta(x-y)\Big) , \nonumber\\
\hspace{-1em} &\hspace{-1em}&\hspace{-1em} \ \Longrightarrow  \int d^dy {\mathcal{M}}_{\Psi|c}(x,y)\big| \chi^1_c(y) \rangle_{s} =    \eta_0\overline{\eta}^+_1\eta_1\Big( l_0 - \beta l_1^+l_1  -\frac{2\beta}{2s-4+ d} l_1l^+_{11}l_1
 \Big)\big| \chi^1_c(x) \rangle_{s}=0 \label{invpsicondn1}
\end{eqnarray}
where the following values are the only non vanishing ones: $\big| \chi^1_c(x) \rangle_{s}=\mathcal{P}^+_1\big| \xi (a^+,y)\rangle_{s-1}$,  $(gh_H$, $gh_{tot})\big| \chi^1_c(x) \rangle = (-1,-1)$.    The non-minimal  Fadde\-ev-Popov operator ${M}_{\Psi|c} (x,y)$, (for $\beta \ne 0$ and  $\mathcal{P}^+_1\hspace{-0.1em}\overline{\mathcal{P}}_1\hspace{-0.1em}\mathcal{P}_0\hspace{-0.1em}\mathcal{M}_{\Psi|c}(x,y)$ =  $\imath{M}_{\Psi|c} (x,y)$) acquires the tensor form
\begin{eqnarray} && \hspace{-0.5em}\int d^dy {M}_{\Psi|c}(a^{(+)};x,y)\big| \xi (a^+,y)\rangle_{s-1} =  \Big( l_0 - \beta l_1^+l_1  -\frac{2\beta}{2s-4+ d} l_1l^+_{11}l_1
 \Big)\big| \xi (a^+,x)\rangle_{s-1},       \label{invpsicond2}\\
 && \hspace{-0.5em}
 {M}_{\Psi|c}{}^{(\nu)_{s-1}}_{(\mu)_{s-1}}(x)\  =\  \partial^2  \delta^{(\nu)_{s-1}}_{(\mu)_{s-1}} +K^{(\nu)_{s-1}}_{(\mu)_{s-1}}\big(\partial, \beta \big)  \  \mathrm{from} \    \left({M}_{\Psi|c}(x)\xi(x)\right) ^{(\mu)_{s-1}} \prod_{i=1}^{s-1}(a^+)_{(\mu)_{i}}|0\rangle,  \label{invpsicond3} \\
   &&
   \hspace{-0.5em}
   K^{(\nu)_{s-1}}_{(\mu)_{s-1}}\big(\partial, \beta \big)  \hspace{-0.15em}=    \hspace{-0.15em} \frac{\beta}{2s-4+ d}\eta^{\{\nu_{s-2}\nu_{s-1}} \partial_{\mu_{s-1}} \partial_{\mu_{s-2}} \delta^{(\nu)_{s-3}\}}_{(\mu)_{s-3}} - \hspace{-0.15em}  \beta\frac{s-3+ (d/2)}{s-2+ (d/2)}  \partial_{\mu_{s-1}}\partial^{\{\nu_{s-1}}
  \delta^{(\nu)_{s-2}\}}_{(\mu)_{s-2}} , \nonumber \\
 &&
 \hspace{-0.5em}
    \mathrm{for} \
   \delta^{(\nu)_{s-k}}_{(\mu)_{s-k}} = \delta^{\{\nu_{1}}_{\mu_{1}}\ldots \delta^{\nu_{s-k}\}}_{\mu_{s-k}}=\frac{1}{(s-k)!}\left[\delta^{\nu_{1}}_{\mu_{1}}\ldots \delta^{\nu_{s-k}}_{\mu_{s-k}}+ cycl.perm.(\nu_{1},...,\nu_{s-k}) \right]\label{invpsicond4}
\end{eqnarray}
 Let us note, in the first place, that we have omitted in
(\ref{invpsicondn1}) the term $\overline{\eta}^+_1\eta_1^+(l_1)^2$
and also its dual $\eta_1\overline{\eta}_1(l^+_1)^2$, since it
vanishes as one estimates the scalar product for the
ghost-antighost term $\int d\eta_0 {}_{s}\big(\langle\overline{C}
\big| {\mathcal{M}}_{\Psi|c}(x,y)\big| C \rangle_{s}+h.c.\big)$.
Secondly, the variational derivatives for a fixed
$\eta_0$,\newline  ${\overleftarrow{\delta}_{\eta_0}}/{\delta
\big| \chi^0_c(y) \rangle_{s}}$,
${\overrightarrow{\delta}_{\eta_0}}/{\delta{}_{s}\langle\chi^{0}_{c}(y)\big|}$
are calculated in accordance with the rules (\ref{transf1}) of the
superfield BRST--BV quantization \cite{LMR, GMR}.

We can now determine the quantum BRST--BV action $S^{\Psi}_{0|s}
\equiv S^{\Psi}_{0|s}\big(\chi^{0}_{\mathrm{tot}|c}\big)$ as a
shift of the vector $\big| \chi^{ 0}_{\mathrm{tot}|c}\rangle_{s}$
by a variational derivative of the gauge-fermion functional:
\begin{eqnarray}
&\hspace{-1em}&\hspace{-1em} \big| \chi^{ 0}_{\mathrm{tot}|c}\rangle_{s} \to \big| \chi^{\Psi{} 0}_{\mathrm{tot}|c}\rangle_{s} = \big| \chi^{ 0}_{\mathrm{tot}|c}\rangle_{s}+ \frac{\overrightarrow{\delta}}{\delta  {}_{s}\langle\chi^{0}_{\mathrm{tot}|c}|}\Psi = \Big\{1+ \eta_0 \frac{\overrightarrow{\delta}_{\eta_0}}{\delta  {}_{s}\langle\chi^{0}_{\mathrm{tot}|c}|}{}_{s}\langle\chi^{0}_{\mathrm{tot}|c}|\Psi_H\Big\} \big| \chi^{ 0}_{\mathrm{tot}|c}\rangle_{s} \label{chi-shift}\\
&\hspace{-1em}&\hspace{-1em} \Longrightarrow  \big( |\chi^{\Psi{}0}_{\mathrm{f}|c}\rangle_{s},\, |\chi^{\Psi{}0}_{\mathrm{af}|c}\rangle_{s} \big) \  = \ \big( |\chi^{0}_{\mathrm{f}|c}\rangle_{s},\, |\chi^{0}_{\mathrm{af}|c}\rangle_{s}  +  \eta_0 \Psi_H\big| \chi^{ 0}_{\mathrm{f}|c}\rangle_{s} \big) , \label{chi-shift1}
\end{eqnarray}
where the antifield components are the only ones that change under
the notation  $\Delta_\Psi |A\rangle \equiv |A^\Psi\rangle - |A\rangle$,
 \begin{align}& \Delta_\Psi  |\chi^{0*}_{c}\rangle_{s} =  -\eta_0 \widehat{\chi}{}^+\big|\overline{C}(a^+)\rangle_{s-1}, && \Delta_\Psi  \big|\overline{C}{}^{*}\big(\overline{\eta}{}_1^+, a^+\big)\rangle_{s}  =  -\eta_0\overline{\eta}{}_1^+\widehat{\chi}\big(\big| \chi^{0}_{c} \rangle_{s}+\big|b(\lambda^+_1)\rangle_{s}\big)  , \label{chi-shiftcomp1}
 \\
 & \Delta_\Psi  |C^{*}\rangle_s = 0 , &&  \Delta_\Psi  |b^{*}(\pi_1^+, a^+)\rangle_{s} =  - (\xi/2)\eta_0 \pi_1^+  \big|\overline{C}(a^+)\rangle_{s-1} . \label{chi-shiftcomp2}
\end{align}
Extended by antifields and usual $\mathcal{S}^{\Psi}_{0|s}$ (for $|\chi^{0}_{\mathrm{af}|c}\rangle = 0$) quantum actions read
\begin{eqnarray}
&&   S^{\Psi}_{0|s}  =    S_{0|s}\big(\chi^{\Psi{}0}_{\mathrm{tot}|c}\big) =   S_{0|s}\big(  \chi^{{}0}_{\mathrm{f}|c}, \chi^{\Psi{}0}_{\mathrm{af}|c}\big)   \ = \  \int d \eta_0 \; {}_{s}\langle \chi^{\Psi{} 0}_{\mathrm{tot}|c}
\big| Q_{c|\mathrm{tot}}\big| \chi^{\Psi{} 0}_{\mathrm{tot}|c}\rangle_{s},
  \label{quantactext}\\
  && \mathcal{S}^{\Psi}_{0|s} = S_{0|s}\big(\widetilde{\chi}{}^{\Psi{}0}_{\mathrm{tot}|c}\big)\ = \ \int d \eta_0 \; {}_{s}\langle \widetilde{\chi}{}^{\Psi{} 0}_{\mathrm{tot}|c}
\big| Q_{c|\mathrm{tot}}\big| \widetilde{\chi}{}^{\Psi{} 0}_{\mathrm{tot}|c}\rangle_{s}, \ \mathrm{for} \  \big| \widetilde{\chi}{}^{\Psi{} 0}_{\mathrm{tot}|c}\rangle \equiv  \big| \chi^{\Psi{} 0}_{\mathrm{tot}|c}\rangle\vert_{ (|\chi^{0}_{\mathrm{af}|c}\rangle = 0)}. \label{quantact1}
\end{eqnarray}
By construction, the functional  $S^{\Psi}_{0|s}$ presented  in the $\eta_0$-independent, ghost-independent and tensor forms
\begin{eqnarray}
 &&   S^{\Psi}_{0|s}   =
\mathcal{S}_{C|s} +
  \biggr(\left({}_{s}\langle S^{*}_c \big|, {}_{s}\langle B^{*}_c \big|  - {}_{s-1}\langle \overline{C} \big| \widehat{\chi} \right)\left(\begin{array}{cc}
  l_0 & - \Delta Q_c\\
 -\Delta Q_c & \eta_1^+\eta_1            \end{array}\right)
       \left(  \begin{array}{c}\big|C\rangle_s \\  0   \end{array} \right)  \label{SqBVfin1}\\
 &&   \phantom{S^{\Psi}_{0|s}=} - \Big({}_{s}\langle \overline{C}{}^*(\overline{\eta}_1,a)\big| -  \big({}_s\langle\chi^{0}_{c}| + {}_s\langle b(\lambda_1)|\big)\widehat{\chi}{}^+\overline{\eta}{}_1\Big) \overline{ \mathcal{P}}{}^+_1\pi_1
| b(\lambda^+_1,a^+) \rangle_{s}+h.c. \biggr) \nonumber \\
&&  \phantom{S_{C|s}}=
{S}_{0|s}+ \Big({}_{s-1}\langle \overline{C} \big| \widehat{\chi}l_1^+ |C\rangle_{s-1} +  \big({}_s\langle\chi^{0}_{c}| + {}_s\langle b(\lambda_1)|\big)\widehat{\chi}{}^+| b(a^+) \rangle_{s-1}+h.c\Big)
 \label{SqBVfin2}\\
 && \phantom{S_{C|s}}  =  {S}_{0|s} +   2 \frac{(-1)^s}{s!} \int d^dx \ s \bigg\{ \big(\overline{C}{M}_{\Psi|c}   C\big)_{s-1} \hspace{-0.15em}+\Big(\partial^{\mu_s} \phi_{(\mu)_{s}}  + 2(s-1)\Big[(1+\beta)\partial_{\mu_s-1}\delta^\rho_{\mu_{s-2}} \nonumber\\
&& \phantom{\Psi_{F|s}}+ \frac{2\beta }{2s-4+ d}\big[2\partial_{\mu_s-1}\delta^\rho_{\mu_{s-2}} +  (s-2) \eta_{\mu_{s-1}\mu_{s-2}}\partial^{\rho}\big] \Big]\phi_{2(\mu)_{s-3}\rho}+ \frac{\xi}{2}b_{(\mu)_{s-1}}\Big) b^{(\mu)_{s-1}}\bigg\}\label{SqBVfin4}
\end{eqnarray}
  satisfies, once again,  the master equation (\ref{antibracketdef}) in the total field-antifield space with the respective antibracket. Note, that we have used in (\ref{SqBVfin4})  a notation  for the ghost Faddeev-Popov term $ \big(\overline{C}{M}_{\Psi|c}   C\big)_{s-1} \equiv$   $\overline{C}_{(\nu)_{s-1}}{M}_{\Psi|c}{}^{(\nu)_{s-1}}_{(\mu)_{s-1}}(x)   C^{(\mu)_{s-1}}$.  The quantum
extended actions $S^{\Psi}_{F|s}$, for instance, in the
single-field (Fronsdal) form for the tensor representation are
obtained from (\ref{SqBVfin4}) by setting $\phi_1^* = \phi_2^*=0$
and expressing $\phi_1$ from the algebraic equation of motion in
terms of $\phi, \phi_2$, as well as $ \phi_2^{(\mu)_{s-2}}=
(1/2)\phi^{(\mu)_{s-2}\mu}{}_{\mu}$. The  quantum action
$\mathcal{S}^{\Psi}_{0|s}$ has a standard structure composed by
the classical, ghost and gauge-fixed parts,
$\mathcal{S}^{\Psi}_{0|s} =
\mathcal{S}_0+S_{\mathrm{gh}}+S_{\mathrm{gf}}$, and coincides with
the known quantum action in the tensor form, modulo the common
factor $2 {(-1)^s}/{s!}$.  The actions  $S^{\Psi}_{0|s}$ and
$\mathcal{S}^{\Psi}_{0|s}$ are both invariant under the BRST
transformations (\ref{totbrst0}) and are also non-degenerate in
the total configuration space of $\phi^A$ (however with resolved traceless constraints), thus providing a naive
definition for the extended  ($Z_0\left[ J^0,\phi^*  \right]$) and
usual ($\chi^{0}_{\mathrm{af}|c} = 0$) generating functionals of
Green's functions, with a new \emph{generalized vector of external
sources}  ${}_s\langle J^{0}_{\mathrm{f}|c}\big|$ ($\big|J
^{0}_{\mathrm{f}|c}\rangle_s$)
 \begin{eqnarray}
&\hspace{-1ex}& \hspace{-1ex} {}_s\langle J^0_{\mathrm{f}|c}\big|  =  {}_s\langle J^0_c\big|  + \langle 0|   J^C_{s-1}(a)\eta_1\eta_0  + \langle J^{\overline{C}}_{s-1}(a)\big|\overline{\eta}_1\eta_0 +
\langle 0| J^{b}_{s-1}(a)\lambda_1 \eta_0    \label{Jconfvbra}
 \\
&\hspace{-1ex}& \hspace{-1ex} {}_s\langle J^0_c\big|  =   \Big({}_s\langle J(a)\big|   +  {}_{s-2}\langle J_2(a)\big| \eta_1\mathcal{P}_1\Big)\eta_0+ {}_{s-1}\langle J_1(a)\big| \eta_1 , \ \ \mathrm{for} \ \ (\epsilon, gh_{tot})\langle J^0_{\mathrm{f}|c}\big| = (1, 1)  \label{Jconfvbra2}
\end{eqnarray}
 to the generalized  field vector  $\big|\chi^{0}_{\mathrm{f}|c}\rangle_s$  (${}_s\langle \chi^{0}_{\mathrm{f}|c}\big|$). These vectors (with appropriate complex conjugation rules for its tensor components) contains the usual sources $|J_k\rangle_{s-k}, ({}_{s-k}\langle J_k|)$ for the field vectors ${}_{s-k}\langle \phi_{k}|$  ($|\phi_{k}\rangle_{s-k}$) for $k=0,1,2$. We determine the naive functional $Z_0\left[ J^0,\phi^*  \right] $  in the form
\begin{equation}
Z_0\left[ J^0,\phi^*  \right] \ = \  \int d \chi^{0}_{\mathrm{f}|c} \exp \bigg\{\frac{\imath}{\hbar} \left[ S^{\Psi}_{0|s}\big(\chi^{0}_{\mathrm{tot}|c}\big)  +  \int d\eta_0 \Big({}_s\langle J^{0}_{\mathrm{f}|c}| \chi^{0}_{\mathrm{f}|c}\rangle_s + {}_s\langle \chi^{0}_{\mathrm{f}|c}| J^{0}_{\mathrm{f}|c}\rangle_s \Big)\right]\bigg\} \label{UVnm}
 \end{equation}
with the measure  $d \chi^{0}_{\mathrm{f}|c} =  \prod_x   d \phi^{(\mu)_s}(x)d \phi_1^{(\mu)_{s-1}}(x)d \phi^{(\mu)_{s-2}}_2(x)dC^{(\mu)_{s-1}}(x)d\overline{C}{}^{(\mu)_{s-1}}(x)db^{(\mu)_{s-1}}(x)$ determined for triplet, doublet [without $d \phi_1$] and  for single-field [without $d \phi_1d\phi_2$] formulations.

\noindent
\textbf{Remarks:} Without off-shell traceless constraints, the respective quantum actions correspond to reducible representations of the $ISO(1,d-1) $ group with multiple helicities $s, s-2, s-4,...,1(0)$, and the functional integral (\ref{UVnm}) is well defined in the  doublet and triplet formulations.

 In the opposite case, the functionals  $S^{\Psi}_{0|s}$ (\ref{quantactext}) and $\mathcal{S}^{\Psi}_{0|s}$  (\ref{quantact1}) -- in order to serve as an extended quantum action and  a quantum action (allowing for addition of consistent interacting terms) which may help to determine the path integral -- should depend on independent (anti)field variables only. Indeed, the variables  in the entire set of fields $\phi^A$, antifields $\phi^*_A$, and sources $J_A$ of the BV quantization method \cite{BV} should be independent, i.e., having no unresolved  external (algebraic or other) constraints, so as to provide a correct application of the Feynman rules to the corresponding generating functional of Green's functions. Therefore, the constraints
\begin{equation*}
 \mathcal{L}_{11}\left(\big|\chi^{0}_{\mathrm{f}|c}\rangle_s, \, \big|\chi^{0}_{\mathrm{af}|c}\rangle_s,\, | J^{0}_{\mathrm{f}|c}\rangle_s\right) = 0
\end{equation*}
should be explicitly resolved. Thus, in the  singlet-field formulation the (anti)\-fields $\phi^{(*)(\mu)_s}$ and  $C^{(*)(\mu)_{s-1}}$, $\overline{C}{}^{(*)(\mu)_{s-1}}$, $b^{(*)(\mu)_{s-1}}$ are, respectively, double-traceless and traceless.

In the  doublet and triplet formulations, we should use a decomposition of the initial (anti)\-field  $\phi^{(*)(\mu)_s}$
and of  the respective vector $|\phi^{(*)} \rangle_s$ in a sum of two traceless fields (also considered in \cite{BRST-BV3}):
\begin{align}\label{decomp2tr}
& |\phi^{(*)} \rangle_s \ =\  |{\phi}_I^{(*)} \rangle_s + \kappa l_{11}^+ |\phi^{(*)}_{II} \rangle_{s-2},\,  && \kappa \equiv  (s+d/2-2)^{-1} , \\
 & l_{11} \left( |{\phi}_I^{(*)} \rangle_s , |{\phi}_{II}^{(*)} \rangle_{s-2}\right)  \  = \  0, \, && l_{11} |\phi^{(*)} \rangle_s \ = \   |\phi^{(*)}_{II} \rangle_{s-2}, \nonumber
\end{align}
which is valid also for the source  $|J_0 \rangle_s$ (\ref{Jconfvbra2}). Comparison with the resolution of constraints (\ref{resconstrf}), (\ref{constraintstar})  permits the identification
\begin{equation}\label{identc}
\left( |\phi_{II} \rangle, |\phi^{*}_{II} \rangle, |J_{II} \rangle\right)  \ = \  \left(- |\phi_2 \rangle, |\phi^{*}_2  \rangle, |J_{2} \rangle\right) .
\end{equation}
As a result, we may equivalently present the actions ${S}^{\Psi}_{0|s}$, $\mathcal{S}^{\Psi}_{0|s}$ of the  doublet and triplet formulations entirely in terms of traceless (anti)\-fields, as we substitute, instead of  $|\phi^{(*)} \rangle_s$, the sums (\ref{decomp2tr}) of the new $|{\phi}_I^{(*)} \rangle_s$ and old $|\phi^{*}_2 \rangle$ traceless antifields, thus changing the basis of field-antifield configurations, and therefore also the structure of the actions and the measure $d \chi^{0}_{\mathrm{f}|c}$, where,  instead of $ d \phi^{(\mu)_s}(x)$, one substitutes $d \phi^{(\mu)_s}_I(x)$. For instance, the BRST-BFV constrained action having only traceless constraints in the triplet  formulation (\ref{gaugetrip}) takes the form
\begin{eqnarray}
   && \mathcal{S}_{C|s}  =  \left(\langle \phi_I  \big|    \langle \phi_2\big| \langle \phi_1\big|   \right)\left(\begin{array}{ccc}
  l_0 &   -\kappa{l^+_{11}l_0} & -l_1^+ \\
  -\kappa{l_{11}l_0} & -l_0+\kappa^2{l_{11}l_0l^+_{11}} &  l_1+\kappa{l_{11}l_1^+}    \\
        -l_1 & l_1^+ +\kappa{l_1 l^+_{11}}  &  1  \end{array}\right)
       \left(  \begin{array}{c}\big|{\phi}_I\rangle\\ \big|{\phi_2}\rangle \\ \big|{\phi_1}\rangle   \end{array} \right) , \label{tripStra} \\
                \label{gaugetr2}
                &&
    \phantom{\mathcal{S}_{C|s}} \delta \left( \big|\phi_I\rangle_{s} ,      \big|\phi_1\rangle_{s-1},    \big|\phi_2\rangle_{s-2}\right) = \left( l_1^+ ,l_0,l_1\right) |\xi\rangle_{s-1}.
\end{eqnarray}
 The same must be done with the antifield-dependent part of the BRST-BV minimal action (\ref{Sgenfin3}), which takes the form
\begin{equation*}
  \biggl(\left[{}_{s-1}\langle \phi^{*}_1 \big|l_0  - {}_{s}\langle \phi^{*}_I \big|l_1^+ - {}_{s-2}\langle \phi^{*}_2 \big|\left(l_1+\kappa  {l_{11}l_1^+}\right)\right] C_{s-1}(a^+)|0\rangle +  h.c.\biggr),\label{Sgenfin3tr}
\end{equation*}
 and then one should make a shift of the gauge parameter $\beta\to (\beta+1)=\beta^{\prime}$ in the gauge fermion (\ref{solmastereq5}).
\vspace{1ex}

The related Green functions  can be obtained by differentiation with respect to the external  sources, e.g., the 2-point function $ G^{(2)}(a^{(+)}; x,y)$ with the initial TS field $|\phi\rangle_s$ in the single-field form (\ref{SclsrsingleF}) for $J_0(x)\equiv J(x)$, $\xi\ne 0$,
\begin{eqnarray}\label{UVnm2}
  G^{(2)}(a^{(+)}; x,y) &  = & \frac{\overrightarrow{\delta}}{\delta {}_s\langle J(x) |} Z_0\left[ J^0,\phi^*  \right]  \frac{\overleftarrow{\delta}}{\delta | J (y) \rangle{}_s}\Big|_{J^0_c=\phi^*=0}\\
  &  = &    \left[  l_0-l_1^+l_1  -(l_1^+)^2l_{11}
 -l_{11}^+l_1^2  -l_{11}^+(l_0 +  l_1l_1^+) l_{11} + \xi^{-1}\widehat{\chi}^+_{0}\widehat{\chi}_{0}\right]^{-1} \delta(x-y) \nonumber,\\
\widehat{\chi}_{0} & =&   \widehat{\chi}\vert_{\xi=0} = l_1 + \big[(1+\beta)l_1^+ + \frac{2\beta}{2s-4+ d} l_1l^+_{11}\big]l_{11},
 \end{eqnarray}
as one uses the equations (\ref{gaugegen}), (\ref{gaugecomp}). In the Feynman gauge, the Green function takes the minimal form
$G^{(2)}_{(\xi,\beta)=(1,0)}=(  l_0-l_{11}^+l_0
l_{11})^{-1}\delta(x-y)$. Note that the function $G^{(2)}(a^{(+)}; x,y) $ is determined as a Green function by acting  in the space of
double-traceless fields. Due to the BRST trans\-for\-ma\-tions (\ref{totbrst0}) for the integrand in $Z_0^\Psi = Z_0\left[
0,\phi^* \right] $, the Ward Identity for $Z_0$ can be presented as follows:
\begin{equation}\label{WI}
\int d\eta_0 \Big({}_s\langle J^{0}_{\mathrm{f}|c}| \frac{\overrightarrow{\delta}}{\delta {}_s\langle \chi^{0}_{\mathrm{af}|c}|} Z_0\left[ J^0,\phi^*  \right] + Z_0\left[ J^0,\phi^*  \right]\frac{\overleftarrow{\delta}}{\delta  |\chi^{0}_{\mathrm{af}|c}\rangle_s} |J^{0}_{\mathrm{f}|c}\rangle_s  \Big) = 0.
\end{equation}
The gauge independence of $Z^\Psi_0 $: $Z^\Psi_0
=Z_0^{\Psi+\delta\Psi}$ upon an admissible change of the gauge
condition $\Psi\to \Psi+\delta\Psi$ (e.g., by varying
$(\xi,\beta)\to $ $(\xi+\delta \xi,\beta+\delta\beta)$) can be
easily established. Inserting, instead of  the quadratic action
$S^{\Psi}_{0|s}$, the action of an interacting model
$S^{\Psi}_{[1]|s}$ constructed according to the recipe
(\ref{BVFronscacint}) with shifted antifields (\ref{chi-shift1}),
we obtain a non-trivial generating functional $Z\left[ J^0,\phi^*
\right]$ of Green's functions in the BRST-BV formalism which leads
to a non-trivial $S$-matrix.

Note that in a gauge theory of interacting HS fields, e.g., one with helicities $s_1, s_2, s_3$, the  BRST-BV minimal action in ${S}_{F|s_1,s_2,s_3} $ should include the sum of BRST-BV minimal actions ${S}_{F|s_i}$ (\ref{Sgenfin5}) for the free fields $\phi^{(i)}_{(\mu)_{s_i}}$, $i=1,2,3$ and the respective deformation of this sum $S_{\mathrm{int}}$, being consistent with the master equation (\ref{antibracketdef}) and written in a joint field-antifield space:
\begin{eqnarray}
\hspace{-0.3em}  &\hspace{-0.3em}& \hspace{-0.3em}{S}_{F|s_1,s_2,s_3} = \sum_{i=1}^3{S}_{F|s_i}\big(\phi^{(i)}, \phi^{(i)*}, C^{(i)}\big)+ S_{\mathrm{int}}\big(\phi^{[3]}, \phi^{[3]*}, C^{[3]}, C^{[3]*}\big), \  \mathrm{for }\ D^{[3]}=(D^{1}, D^{2}, D^{3}), \nonumber \\
 \hspace{-0.3em}  &\hspace{-0.3em}& \hspace{-0.3em}  S_{\mathrm{int}} = g\int \prod_i d \eta^{i}_0\Big(  \otimes_{j=1}^3{}_{s_j}\langle\chi^{0{}j}_{\mathrm{g}|c}\big|V_{g}\rangle_{s_1,s_2,s_3}+ {}_{s_1,s_2,s_3}\langle V^+_{g} \big| \otimes_{j=1}^3 \big|\chi^{0{}j}_{\mathrm{g}|c}\rangle_{s_j}
\Big)   \label{sintform}
\end{eqnarray}
with operator $\big|V_{g}\rangle_{s_1,s_2,s_3}$ and its hermitian conjugated  ${}_{s_1,s_2,s_3}\langle V^+_{g} \big| $ which determined a cubic vertex to be consistent with traceless constraints $\widehat{L}^{i}_{11}$ for respective vector.
The total  BRST-BV quantum action $S^{\Psi^{[3]}}_{s_1,s_2,s_3}\big(\chi^{0[3]}_{\mathrm{tot}|c}\big)$ should additionally provide the introduction of an admissible gauge condition $\Psi^{i}$ having the form (\ref{solmastereq3}), (\ref{solmastereq3ferm}) for each HS field $\phi^{(i)}_{(\mu)_{s_i}}$,
 which looks as follows:
\begin{equation}\label{QAint}
  S^{\Psi^{[3]}}_{s_1,s_2,s_3}\big(\chi^{0[3]}_{\mathrm{tot}|c}\big) = \sum_{i=1}^3 S^{\Psi^{i}}_{s_i}\big(\chi^{0i}_{\mathrm{tot}|c}\big) + S_{\mathrm{int}}\big(\phi^{[3]}, \phi^{\Psi^{[3]}[3]*}, C^{[3]}, C^{[3]*}\big),
\end{equation}
where the form of $i$'s copy is determined by (\ref{quantactext}) and for 3 copies of antifields  $\phi^{\Psi^{[3]}[3]*}$ (with use of the notation (\ref{sintform})) the respective  shift (\ref{chi-shift})--(\ref{chi-shiftcomp2}) has been made.
As a result, in accordance with the above Remark the functional integral $Z^{[3]} = \exp \{(i/\hbar)W^{[3]}\}$,
\begin{equation}
Z^{[3]}\left[ J^{0[3]},\phi^{[3]*}  \right] \ = \  \int \prod_i d \chi^{0i}_{\mathrm{f}|c} \exp \bigg\{\frac{\imath}{\hbar} \bigg[ S^{\Psi^{[3]}}_{s_1,s_2,s_3}\big(\chi^{0[i]}_{\mathrm{tot}|c}\big)  +  \int d\eta_0 \sum_{i=1}^3\Big({}_{s_i}\langle J^{0i}_{\mathrm{f}|c}| \chi^{0i}_{\mathrm{f}|c}\rangle_{s_i} + h.c. \Big)\bigg]\bigg\}, \label{UVnmreal}
 \end{equation}
correctly determines  the generating functionals $Z^{[3]}$, $W^{[3]}$ for general and  connected Green's functions, and therefore also determines the effective action. The Ward identity (\ref{WI}) deduced for $Z^{[3]}$, $W^{[3]}$ has a similar form (for $Y = \{Z,W\}$)
\begin{equation}\label{WIint}
 \sum_{i=1}^3\int d \eta^{i}_0\Big({}_{s_i}\langle J^{0i}_{\mathrm{f}|c}| \frac{\overrightarrow{\delta}}{\delta {}_s\langle \chi^{0i}_{\mathrm{af}|c}|} Y^{[3]}\left[ J^{0[3]},\phi^{[3]*}  \right] + Y^{[3]}\left[ J^{0[3]},\phi^{[3]*}  \right]\frac{\overleftarrow{\delta}}{\delta  |\chi^{0i}_{\mathrm{af}|c}\rangle_{s_i}} |J^{0i}_{\mathrm{f}|c}\rangle_{s_i}  \Big) = 0.
\end{equation}
They follow from the BRST symmetry transformations for the integrand in (\ref{UVnmreal}) for vanishing sources of the form:
\begin{eqnarray}
 \hspace{-0.5ex} \delta_B |\chi^{0i}_{\mathrm{f}|c}(x) \rangle_{s_i}
\hspace{-0.5ex}& \hspace{-1ex}=  & \hspace{-1ex} \mu \frac{\overrightarrow{\delta}}{\delta {}_{s_i}\langle \chi^{0i}_{\mathrm{af}|c}(x) \big|}S^{\Psi^{[3]}}_{s_1,s_2,s_3}\big(\chi^{0[3]}_{\mathrm{tot}|c}\big)  ,  \label{totbrstint}
\end{eqnarray}
with account for the BRST invariance for the quantum action (in assumption of local form of the interacting summand $S_{\mathrm{int}}$, and with additional relation for non-local $S_{\mathrm{int}}$).
\section{Conclusion}

\label{Concl} 
We have extended the constrained BRST-BFV and BRST-BV methods for constructing the irreducible gauge-invariant Lagrangian $\mathcal{S}_{C|s}= \int d\eta_0 {}_s\langle\chi^0_c| Q_c|\chi^0_c\rangle_s$  (\ref{PhysStatetot})  and minimal BRST-BV   $S_{C|s}  =  \int d \eta_0 \; {}_{s}\langle \chi^0_{\mathrm{g}|c}
| Q_{c}| \chi^0_{\mathrm{g}|c} \rangle_{s}$ (\ref{Sgenfin1}) actions for massless totally-symmetric tensor field of helicity $s$  up to the non-minimal BRST--BV   method in order to obtain the quantum  BV action, ,$\mathcal{S}^{\Psi}_{0|s} =  \int d \eta_0 \; {}_{s}\langle \widetilde{\chi}{}^{\Psi{} 0}_{\mathrm{tot}|c}
\big| Q_{c|\mathrm{tot}}\big| \widetilde{\chi}{}^{\Psi{} 0}_{\mathrm{tot}|c}\rangle_{s}$  (\ref{quantact1})  and the generating functional of Green's functions (\ref{UVnm}) explicitly in terms of the appropriate Fock space vectors.  These vectors contain as their component functions the
whole set of the fields in the respective triplet, doublet and
single-field formulations, together with the ghost, antighost and
Nakanishi-Lautrup fields and respective external sources, with the
minimal Hamiltonian BFV ghost oscillators augmented by additional
four
oscillator pairs  $\overline{\eta}{}_1,  \overline{ \mathcal{P}}{}^+_1$; $\overline{\eta}{}^+_1, \overline{ \mathcal{P}}{}_1$;   $\lambda_1, \pi_1^+$; $\lambda^+_1, \pi_1 $  (\ref{nmingras})   from the non-minimal sector.  The latter operators are shown to be
necessary, in the first place, for augmenting the generalized
vector $|\chi _{\mathrm{g}|c}^{0}\rangle _{s}$, the BRST operator
$Q_{c}$, the BRST-extended constraint $\widehat{L}_{11}$ and the
spin
operator $\widehat{\sigma }_{c}$ up to the respective total quantities $\big|%
{\chi }{}_{\mathrm{tot}|c}^{0}\rangle _{s}$, $Q_{c|\mathrm{tot}}$, $\mathcal{%
L}_{11}$ and ${\sigma }_{c|\mathrm{tot}}$. Secondly, they
are used  to formulate
a Lagrangian gauge-fixing fermion functional $\Psi (\chi _{\mathrm{tot}%
|c}^{{}0})$ (\ref{solmastereq3}), with the help of a Hamiltonian
operator gauge-fixing fermion  $\Psi _{H}$ (\ref{solmastereq3ferm}), in fact  as its  kernel.   The gauge-fixing  fermion corresponds to the 2-parametric family of the gauges which extends the case  .of $R_\xi$-gauges.  The non-minimal BV action and the gauge-fixed
quantum action, obtained using a shift of the
antifield components in the total generalized vector $\big|{\chi }{}_{%
\mathrm{tot}|c}^{0}\rangle _{s}$ by a variational derivative of
the gauge-fermion functional, $\big|\widetilde{\chi
}{}_{\mathrm{tot}|c}^{\Psi {}0}\rangle _{s}$ (\ref{chi-shift}),
(\ref{chi-shift1}), satisfy a master equation and are (along with
the integrand of the vacuum generating functional) invariant under
the Lagrangian BRST transformations (\ref{totbrst0}).

We have obtained different representations for the quantum action in the so-called $\eta_0$-independent, ghost-independent and tensor forms for both the triplet and doublet formulations, as well as for the single-field (Fronsdal) formulation, which are naturally deduced from the BRST-BV quantum action $\mathcal{S}^{\Psi}_{0|s}$. The requirement of absence for any external constraint on the entire set of field, antifield, and source variables, so as to provide a correct definition for the functional integral in the BV quantization method, has been implemented in a singlet-field formulation by using a double-traceless initial  (anti)field and traceless remaining (anti)\-fields. In the  doublet and triplet formulations, the representation of a respective field-antifield space with traceless tensor (anti)\-fields has been used according to (\ref{decomp2tr}), (\ref{identc}).

 The suggested non-minimal BRST--BV approach to constructing a
quantum action for free and interacting massless TS HS fields
allows one to formulate the Feynman quantization rules explicitly
in terms of a generating functional of Green's functions
$Z_0\left[ J^0,\phi^*  \right]$ and for interacting HS theory  $Z^{[3]}\left[ J^{0[3]},\phi^{[3]*}  \right]$ (\ref{UVnmreal})  determined using  generalized
vectors of external sources, and also to finalize the concept of
BFV--BV duality between Hamiltonian and Lagrangian quantities in a
way different from that suggested in \cite{GMR}.

Remarkably, all the ingredients required for conventional
quantization according to perturbation theory -- as regards
establishing the gauge-independence of the vacuum functional from
the choice of admissible gauge conditions, as well as deriving the
Ward identity and formulating the Green functions and the
Faddeev--Popov operator in a manifest form -- can be provided
using operations with Fock-space vectors, as has been done earlier
in the superfield Lagrangian quantization \cite{LMR}, \cite{GMR}.

There are many directions for applications and development  of the
suggested approach, such as the quantum action and Feynman rules for a constrained
TS fields of helicity $s$ in anti-de-Sitter backgrounds, as well as for
unconstrained TS integer HS fields in Minkowski spaces.

\vspace{-1ex}

\paragraph{Acknowledgements}
The authors are thank to I.L. Buchbinder, P.M. Lavrov, P.Yu. Mo\-shin and M. Najafizadeh
for remarks and helpful discussions. A.A.R. appreciates the comments made
by conference participants at QFTG'2016 and QFTG'2018, where some of the results
presented in this research were announced. The work of A.A.R. has been
supported by the Program of Fundamental Research under the Russian
Academy of Sciences, 2013--2020.

\end{document}